\documentclass[a4paper,final,fleqn]{cas-sc}

\usepackage[numbers,sort&compress]{natbib}

\usepackage{siunitx}
\usepackage{upgreek}
\usepackage[%
cal=cm,
calscaled=1.0,%
]{mathalpha}

\usepackage{dcolumn}
\usepackage{nicematrix}
\usepackage{derivative} 
\usepackage{mathtools}
\usepackage{algorithm}
\usepackage{algpseudocode}
\algnewcommand\algorithmicinput{\textbf{Input:}}
\algnewcommand\Input{\item[\algorithmicinput]}
\algnewcommand{\LineComment}[1]{\Statex \(\triangleright\) \textit{#1}}
\usepackage{makecell}
\usepackage{multirow}
\usepackage{booktabs}

\DeclareSIUnit\corehour{\text{core-hours}}
\usepackage{hyperref}
\usepackage[nameinlink,capitalize]{cleveref}
\usepackage{enumitem}
\usepackage{lineno}

\newcommand{\mymatht}[1]{\mbox{\ensuremath{#1}}}
\newcommand{\mymathtv}[1]{\ensuremath{#1}}

\newcommand{\myOrdernum}[1]{\ensuremath{\mathcal{O}(\num{#1})}}
\newcommand{\myOrderqty}[2]{\ensuremath{\mathcal{O}(\num{#1})\,\unit{#2}}}

\newcommand{\mysuchthat}{\ensuremath{:}}

\newcommand{\myvect}[1]{\ensuremath{\boldsymbol{\mathrm{#1}}}}

\newcommand{\myabs}[1]{\ensuremath{\lvert#1\rvert}}

\newcommand{\mysum}[2]{\sum\limits_{#1}^{#2}}

\newcommand{\myNat}{\mathbb{N}}

\newcommand{\myReall}[1]{\mathbb{R}_{#1}}
\newcommand{\myNatu}[1]{\mathbb{N}^{#1}}

\newcommand{\myRealul}[2]{\mathbb{R}^{#1}_{#2}}

\newcommand{\mycode}[1]{\texttt{#1}}

\newcommand{\myAmBe}{\text{AmBe}}
\newcommand{\myPuBe}{\text{PuBe}}
\newcommand{\myCf}{\text{Cf-252}}
\newcommand{\myPu}{\text{Pu-240}}
\newcommand{\myDD}{\text{DD}}
\newcommand{\myDT}{\text{DT}}

\newcommand{\mynuclt}[3]{\ensuremath{\cramped{\prescript{#1}{#2}{\mathrm{#3}}}}}
\newcommand{\myCfiso}{\mynuclt{252}{98}{Cf}}

\newcommand{\myCfstr}{\ensuremath{\xi_{\scriptscriptstyle{\myCf}}}}
\newcommand{\myPustr}{\ensuremath{\xi_{\scriptscriptstyle{\myPu}}}}

\newcommand{\myDTstr}{\ensuremath{\xi_{\scriptscriptstyle{\myDT}}}}

\newcommand{\mydispersion}{\ensuremath{\alpha_{\scriptscriptstyle\text{NB}}}}

\begin{document}

\let\WriteBookmarks\relax
\def\floatpagepagefraction{1}
\def\textpagefraction{.001}

\shorttitle{Bayesian evidence adaptive pursuit\ldots}    

\shortauthors{}  

\title [mode = title]{Disentangling mixed neutron fields: multi‑source identification from few detected events}  

\author[1]{David Breitenmoser}[orcid=0000-0003-0339-6592]
\cormark[1]
\ead{dbreiten@umich.edu}
\credit{Conceptualization, Formal analysis, Investigation, Methodology, Project administration, Software, Validation, Visualization, Data curation, Writing -- original draft, Writing -- review and editing}

\author[1]{William Heriot}[orcid=0000-0003-2964-5745]
\credit{Software, Investigation, Writing -- review and editing}

\author[2]{Peter Marleau}[orcid=0009-0001-7202-316X]
\credit{Methodology, Software, Writing -- review and editing}

\author[1]{Shaun D. Clarke}[orcid=0000-0003-2469-3166]
\credit{Conceptualization, Methodology, Software, Writing -- review and editing, Funding acquisition}

\author[1]{Sara A. Pozzi}[orcid=0000-0001-6827-3652]
\credit{Conceptualization, Methodology, Writing -- review and editing, Funding acquisition}

\affiliation[1]{organization={Department of Nuclear Engineering \& Radiological Sciences, University of Michigan},
            addressline={2355 Bonisteel Blvd.}, 
            city={Ann Arbor},
            postcode={MI 48109-2104}, 
            country={\\United States of America}}
\affiliation[2]{organization={Radiation and Nuclear Detection Systems Division, Sandia National Laboratories},
            city={Livermore},
            postcode={CA 94551}, 
            country={United States of America}}

\cortext[1]{Corresponding author}

\begin{abstract}
Identifying neutron-emitting materials is central to nuclear nonproliferation, safeguards, nuclear forensics, and emergency response, yet remains difficult when several sources contribute simultaneously: relevant fission, $(\alpha,\text{n})$, and fusion sources emit broad, strongly overlapping energy distributions, and the associated spectral inversion is severely ill-conditioned. Here we demonstrate quantitative identification of mixed neutron fields directly from scatter-based (recoil) spectroscopy measurements, together with simultaneous estimation of the emission rate of each contributing source and a rigorous statistical confidence level for every candidate source combination. Using a compact \qty{21.6}{\cubic\cm} organic-glass scintillator spectrometer, we correctly identify Cf-252, a deuterium--deuterium (DD) neutron generator, and their mixture with decisive statistical support (\mbox{$>\!4\sigma$}), and further resolve a weak deuterium--tritium contaminant in the nominal DD generator field. High-fidelity Monte Carlo simulations spanning exhaustive single-, two-, and three-source mixtures show that identification requires remarkably little information: between $\mathcal{O}(10^1)$ and $\mathcal{O}(10^6)$ detected recoil events, set primarily by spectral similarity, mixture complexity, and emission-rate imbalance. For the compact spectrometer used here, this corresponds to acquisition times as short as a few minutes. These results substantially extend the operational reach of simple single-volume neutron spectrometers, enabling rapid, quantitative, and confidence-calibrated attribution of complex neutron fields in field-deployable instruments.
\end{abstract}


\maketitle


\section{Introduction}
\label{sec:Introduction}

\noindent The identification of neutron-emitting materials plays a central role in nuclear nonproliferation, safeguards, nuclear forensics, and homeland security \cite{alhamrashdiPassiveGammaRayNeutron2019}. Applications such as the interdiction of illicit special nuclear material (SNM), treaty verification, and radiological emergency response require rapid and reliable discrimination between different neutron-emitting sources.

Among available neutron detection modalities, scatter-based neutron spectroscopy---referred to hereafter as recoil spectroscopy---offers several important advantages for field applications. In contrast to segmented scatter-camera systems or time-of-flight (TOF) instruments, scatter-based spectrometers can operate using comparatively simple single-volume detector configurations while still preserving spectral information relevant for source identification. Previous studies have also shown that recoil spectroscopy can provide improved identification performance at both the event-count and acquisition-time level compared to TOF-based approaches \cite{Breitenmoser2025j}. These characteristics make recoil spectroscopy particularly attractive for portable and operationally deployable neutron identification systems.

Despite these advantages, reliable neutron source identification from measured recoil spectra remains fundamentally challenging. Neutron sources relevant to safeguards and nonproliferation applications typically emit broad continuous energy distributions with substantial spectral similarity. Detector response broadening, finite counting statistics, and environmental neutron moderation further obscure distinguishing spectral features, resulting in a severely ill-conditioned inverse problem. As a result, current neutron identification methodologies are limited to qualitative spectral metrics or proxy observables derived from limited portions of the neutron spectrum \cite{Wallenius2007,Mayer2013,Meng2018,Lepowsky2021,LaMont1998}. Although these methods can provide useful supplementary information, they generally lack the ability to perform robust, quantitative identification of complex source ensembles directly from neutron spectral measurements. Secondary gamma-ray signatures may be absent or strongly attenuated depending on source encapsulation and shielding conditions. Similarly, qualitative spectral metrics and correlation-based methods are typically limited to single-source scenarios and do not naturally provide statistically rigorous measures of confidence.

Recently, we demonstrated that Bayesian full-spectrum inference combined with machine-learning-based forward modeling can overcome these limitations and enable statistically rigorous neutron source identification directly from measured recoil spectra \cite{Breitenmoser2025j}. Although this approach achieved high-confidence identification of single- and two-source ensembles, two important limitations remained that restrict practical scalability and deployment. First, the forward model relied on data-driven machine learning approaches trained using dedicated measurement campaigns, resulting in substantial experimental overhead and limited flexibility when extending the framework to new detector configurations or source classes. Second, source inference was performed using an exhaustive combinatorial search strategy whose computational complexity increased exponentially with the dimensionality of the source ensemble. Although tractable for small numbers of candidate sources, this approach becomes computationally prohibitive for larger or more realistic source spaces.

In this work, we address these limitations by introducing a Bayesian evidence adaptive pursuit algorithm for neutron source identification. The proposed methodology replaces data-driven forward modeling with physics-based template generation, substantially reducing dependence on large experimental training datasets while improving generalizability across detector configurations and source classes. Simultaneously, the adaptive pursuit framework enables efficient exploration of high-dimensional source spaces without exhaustive combinatorial enumeration, allowing scalable inference of complex multi-source ensembles. The performance of the proposed approach is evaluated using dedicated laboratory measurements involving spontaneous fission and fusion neutron sources acquired using recoil spectroscopy.

\section{Methods}
\label{sec:Methods}

\subsection{Bayesian model comparison}
\label{sub:BayesMethod}

\noindent Following our previous work \cite{Breitenmoser2025j}, neutron source identification was formulated as a Bayesian model comparison problem in which competing neutron source ensembles are evaluated according to their posterior support conditioned on the measured recoil spectrum \cite{trotta2008a,vonToussaint2011a}. For a measured spectrum \mymatht{\myvect{y}\in\myNatu{M}} with \mymatht{M\in\myNat} spectral channels, the expected detector response was modeled as a linear superposition of source-specific recoil templates,

\begin{equation}
\mathcal{M}(\myvect{\xi};\mathcal{S})
=
\mysum{i\in \mathcal{S}}{} \xi_i \myvect{\uppsi}_i t,
\label{eq:forwardmodelBEAP}
\end{equation}

\noindent where \mymatht{t} denotes the measurement live time, \mymatht{\myvect{\uppsi}_i\in\myRealul{M}{+}} denotes recoil template, i.e., the source-strength-normalized recoil response associated with the $i$-th candidate neutron source, \mymatht{\xi_i\in\myReall{+}} represents the corresponding neutron emission rate, and \mymatht{\mathcal{S}\subseteq\{1,\dots,N\}} defines the active source subset under consideration. Collectively, the template ensemble forms the live-time-scaled sensing matrix \mymathtv{\myvect{M} \coloneqq [ \myvect{\uppsi}_{1},\allowbreak \myvect{\uppsi}_{2},\cdots,\allowbreak \myvect{\uppsi}_{N}]\;t \in \myRealul{M\times N}{+}} containing the recoil response templates of all $N$ candidate neutron sources.

For a given source subset \mymatht{\mathcal{S}}, Bayesian inference was performed over the parameter vector \mymatht{\myvect{\uptheta}_\mathcal{S}}, consisting of the neutron emission rates and nuisance parameters associated with the corresponding forward model. Bayesian inference provides a robust probabilistic framework for evaluating competing hypotheses in ill-conditioned inverse problems, and has been successfully applied in domains such as gravitational-wave astronomy \cite{Veitch2015,Ashton2019,Smith2020} and exoplanet atmospheric retrieval \cite{MacDonald2017,Pinhas2018}. The statistical support for each source subset was quantified through the Bayesian evidence \cite{trotta2008a},

\begin{equation}
\mathcal{Z}_\mathcal{S}
=
\int_{\Theta_\mathcal{S}}
\mathcal{L}(\myvect{\uptheta}_\mathcal{S};\myvect{y},\myvect{M}_\mathcal{S})
\,p(\myvect{\uptheta}_\mathcal{S})
\,\mathrm{d}\myvect{\uptheta}_\mathcal{S},
\label{eq:evidenceBEAP}
\end{equation}

\noindent where $\mathcal{L}(\myvect{\uptheta}_\mathcal{S};\myvect{y},\myvect{M}_\mathcal{S})$ and $p(\myvect{\uptheta}_\mathcal{S})$ denote the likelihood and parameter prior associated with the source subset \mymatht{\mathcal{S}}, respectively. Assuming noncommittal model priors, the relative support between competing source ensembles is determined directly through the ratio of their evidence values \mymatht{\mathcal{B}_{ij} =\mathcal{Z}_i/\mathcal{Z}_j} corresponding to the Bayes factor between the models $\mathcal{M}_i$ and $\mathcal{M}_j$ \cite{vonToussaint2011a}.

As in our previous work \cite{Breitenmoser2025j}, neutron recoil spectra were modeled within a Bayesian full-spectrum inference framework using a negative-binomial likelihood \cite{Hunnefeld2022a,Salinas2020a,Lloyd-Smith2007a} to account for count overdispersion beyond intrinsic Poisson statistics \cite{Praszalowicz2011,Tezlaf2023,Perez2021,Fry2013,Hurtado-Gil2017,Hameeda2021}. The inferred parameter vector \mymatht{\myvect{\uptheta}} comprised the neutron emission rates (\mymatht{\xi\in\myReall{+}}) and the likelihood dispersion parameter (\mymatht{{\mydispersion}\in\myReall{+}}). To enforce physical admissibility while avoiding overly restrictive prior assumptions, weakly informative and statistically independent marginal parameter priors were assigned to the active source rates and the dispersion parameter. For a source subset \mymatht{\mathcal{S}} containing \mymatht{N} active source components, the joint parameter prior was factorized as
\mymathtv{
\pi\left(\myvect{\uptheta}\mid\mathcal{S}\right)
=
\pi\left(\mydispersion\mid\mathcal{S}\right)
\prod_{i=1}^{N}
\pi\left(\xi_i\mid\mathcal{S}\right).
}
Each neutron source emission rate was modeled using a truncated normal prior on the positive real domain with mean and standard deviation \qty{1d8}{\per\s}, while the likelihood dispersion parameter was assigned a positive truncated normal prior with unit mean and unit standard deviation. Complete likelihood and prior definitions are provided in Ref.~\cite{Breitenmoser2025j}.

Bayesian evidences and posterior distributions were evaluated numerically using the nested sampling code \mycode{dynesty} (version~\mycode{3.0.0}) \cite{Speagle2020a} following the inference protocol detailed in Ref.~\cite{Breitenmoser2025j}. The Bayesian inference was performed independently for each candidate source subset \mymatht{\mathcal{S}}, yielding posterior samples \mymatht{\hat{\Theta}_\mathcal{S} \sim p(\myvect{\uptheta}_\mathcal{S} \mid \myvect{y},\mathcal{M}_\mathcal{S})} together with the corresponding Bayesian evidence values \mymatht{\mathcal{Z}_\mathcal{S}} required for model comparison and neutron source identification.

\subsection{Bayesian evidence adaptive pursuit}
\label{sub:BEAP}

\noindent While the inference framework introduced in \cref{sub:BayesMethod} enables statistically rigorous source identification, direct Bayesian model comparison becomes computationally prohibitive for large candidate source spaces. For a source library containing $N$ candidate neutron sources, exhaustive evaluation requires computation of the evidence for all admissible source subsets \mymatht{\Xi=\{\mathcal{S} \subseteq \{1,\dots,N\} \mysuchthat \mathcal{S}\neq\varnothing\}}, resulting in a total of \mymathtv{\myabs{\Xi} = 2^N-1} possible source combinations. Consequently, the computational complexity of exhaustive Bayesian model comparison scales exponentially with increasing source-library dimensionality. This rapidly renders brute-force inference intractable for realistic neutron identification scenarios.

To address this limitation, we introduce the Bayesian Evidence Adaptive Pursuit (BEAP) algorithm, an adaptive model-space exploration strategy designed to efficiently identify high-evidence source ensembles without exhaustive combinatorial enumeration. This methodology iteratively constructs higher-order source mixtures using only source indices associated with statistically competitive lower-order models. A comprehensive pseudo-code of the proposed algorithm is provided in \cref{alg:beap}.

\begin{algorithm}
\caption{Bayesian Evidence Adaptive Pursuit (BEAP)}\label{alg:beap}
\begin{algorithmic}[1]

\State \textbf{Input:}  

Data $\myvect{y}\in\myNatu{M}$

Sensing matrix $\myvect{M}\in\myRealul{M \times N}{+}$

Likelihood $\mathcal{L}(\myvect{\uptheta};\myvect{y} , \myvect{M})$

Parameter prior $p(\myvect{\uptheta})$

Source threshold $\{ D \in \myNat : 1 \leq D \leq N \}$ 

Mixture-retention threshold $\{ B \in \myNat : 1 \leq B \leq N \}$

\State \textbf{initialize:}  

$\mathcal{U} \gets \varnothing$ 

$\Lambda_0 \gets \{ i \in \myNat : 1 \leq i \leq N \}$

$(\log \mathcal{Z}_{\text{max}},\hat{\Theta}_{\text{best}},S_{\text{best}}) \gets (-\infty,\varnothing,\varnothing)$

\For{$k = 1$ to $D$}
    \State $l \gets 0$

    \For{\textbf{all} $\mathcal{S} \in \binom{\Lambda_{k-1}}{k}$}
        \State $l \gets l+1$

        \State $(\log \mathcal{Z}_l, \hat{\Theta}_l) \gets 
        \texttt{EvidenceEstimator}\left(\myvect{y},\mathcal{L}(;\myvect{y} \mid \myvect{\uptheta}_\mathcal{S}, \myvect{M}_\mathcal{S}),p(\myvect{\uptheta}_\mathcal{S})\right)$
        
        \State $S_l \gets \mathcal{S}$
        \State $\mathcal{U} \gets \mathcal{U} \cup \{(\log \mathcal{Z}_l, \hat{\Theta}_l, S_l)\}$
        
        \If{$\log \mathcal{Z}_l > \log \mathcal{Z}_{\text{max}}$}
            \State $(\log \mathcal{Z}_{\text{max}}, \hat{\Theta}_{\text{best}}, S_{\text{best}}) \gets (\log \mathcal{Z}_l,  \hat{\Theta}_l, S_l)$
        \EndIf
    \EndFor
    
    \State $\sigma \gets \{ i \in \mathbb{N} \mid 1 \le i \le l \}, \;\; \forall i,j \in \mathbb{N}, 1 \le i < j \le l : \log \mathcal{Z}_{\sigma(i)} \ge \log \mathcal{Z}_{\sigma(j)}$

    \State $ m \gets \min(B,l)$
    
    \While {$m+1 \leq l \; \wedge \; \log \mathcal{Z}_{\sigma(m+1)}\geq\frac{(m+1)\log \mathcal{Z}_{\sigma(m)}-\log \mathcal{Z}_{\sigma(1)}}{m}$}
        \State $m \gets m + 1$
    \EndWhile
    
    \State $\Lambda_k \gets \bigcup_{i=1}^{m} S_{\sigma{(i)}}$

\EndFor 
\State \textbf{return} $(\log \mathcal{Z}_{\text{max}}, \hat{\Theta}_{\text{best}}, S_{\text{best}}), \mathcal{U}$

\end{algorithmic}
\end{algorithm}

Initially, the candidate source index set is defined as \mymatht{\Lambda_0=\{i\in\mathbb{N}:1\leq i\leq N\}}, corresponding to the full source library. At iteration order $k$, admissible source subsets are defined as $\mathcal{S} \in \binom{\Lambda_{k-1}}{k}$, i.e., the set of all $k$-element subsets of the candidate source index set retained from the previous iteration.  For each subset \mymatht{\mathcal{S}}, we compute the corresponding evidence $\mathcal{Z}_\mathcal{S}$ and posterior parameter estimates $\hat{\Theta}_\mathcal{S}$. Following the evaluation of all candidate mixtures at iteration order $k$, the subsets are ranked according to descending evidence, \mymathtv{\log\mathcal{Z}_{\sigma(1)}\geq\log\mathcal{Z}_{\sigma(2)}\geq\dots\geq\log\mathcal{Z}_{\sigma(l)}}, where $\sigma(i)$ denotes the index permutation corresponding to the ordered evidence values and $l$ is the total number of evaluated subsets at the current iteration level.

The key feature of BEAP enabling efficient exploration of large source libraries is the adaptive reduction of the candidate source space using an Occam-window source-retention procedure. Rather than following a purely greedy strategy that retains only the highest-evidence source mixture and discards all alternatives, BEAP preserves source indices from a broader ensemble of statistically competitive mixtures, thereby preventing the premature elimination of source components whose statistical evidence emerges only in higher-order source combinations. To initialize the retention window, the algorithm first defines \mymatht{m=\min(B,l)}, where $B\in\myNat$ represents the minimum number of candidate mixtures retained at each iteration level and $l$ denotes the total number of evaluated mixtures at the current iteration order. The retention window is subsequently expanded adaptively according to the average logarithmic evidence separation between the maximum-evidence model and the currently retained mixtures. Defining the average log-evidence spacing within the current retention window as

\begin{equation}
\Delta_m
=
\frac{
\log \mathcal{Z}_{\sigma(1)}
-
\log \mathcal{Z}_{\sigma(m)}
}{m},
\label{eq:deltam}
\end{equation}

\noindent an additional candidate mixture is retained if its log-evidence does not decrease more rapidly than the average evidence decay within the current window, i.e.,

\begin{equation}
\log \mathcal{Z}_{\sigma(m+1)}
\geq
\log \mathcal{Z}_{\sigma(m)}
-
\Delta_m.
\label{eq:occamcriterion1}
\end{equation}

\noindent Substituting \cref{eq:deltam} into \cref{eq:occamcriterion1} yields the adaptive Occam-window criterion

\begin{equation}
\log \mathcal{Z}_{\sigma(m+1)}
\geq
\frac{
(m+1)\log \mathcal{Z}_{\sigma(m)}
-
\log \mathcal{Z}_{\sigma(1)}
}{m}.
\label{eq:occamwindow}
\end{equation}

\noindent This adaptive criterion defines an evidence-based Occam window that includes not only the highest-ranked mixtures, but also additional models whose evidence remains sufficiently competitive relative to the current retained ensemble. Intuitively, the procedure balances model parsimony against robustness by preferentially retaining source combinations that remain statistically plausible under Bayesian model comparison while adaptively suppressing mixtures with rapidly decreasing evidence support. After completion of the adaptive retention step, the candidate source index set for the subsequent iteration is defined as the union of source indices contained within the retained high-evidence mixtures. Higher-order candidate ensembles are then constructed exclusively from this reduced source space up to a user defined source threshold \mymathtv{\max k = D \leq N}

The computational complexity of BEAP is governed by the source threshold $D$ and the mixture-retention threshold $B$. In the limiting case where all candidate mixtures are retained at every iteration level (\mymatht{B \geq l}), the algorithm evaluates \mymatht{\sum_{k=1}^{D}\binom{N}{k}} candidate source combinations, corresponding to all admissible mixtures up to order $D$. For $D=N$, this recovers the exhaustive search complexity \mymathtv{2^N-1}. In contrast, for the most aggressive pruning configuration (\mymatht{B=1}), the algorithm approaches greedy search behavior by preferentially propagating only the dominant evidence-supported source combinations between successive iteration levels up to iteration level \mymatht{D}. Intermediate values of $B$ provide a continuous trade-off between computational efficiency and robustness against premature source elimination, while the adaptive Occam-window expansion further allows the retained search space to grow automatically in regions of comparable Bayesian evidence. In this work, we adopted a conservative configuration by setting the source threshold to the maximum number of sources expected in the scene (\mymatht{D=3}) and the mixture-retention threshold to \mymatht{B=15}. This choice ensures that all single-, two-, and three-source hypotheses are explored. For a source library of size $N=10$, the adopted pruning configuration reduces the candidate space from \mymatht{1023} to \mymatht{175} evaluated mixtures, corresponding to an \qty{83}{\percent} reduction in the search space. 

The practical computational cost of BEAP is dominated by the evidence estimator employed in \cref{alg:beap}. The computation time for a single model evidence evaluation, denoted by \mymatht{\Delta t_{\mathcal{Z}}}, may vary by several orders of magnitude depending on the adopted inference method, ranging from sub-millisecond evaluations for simple deterministic approximations such as the Bayesian information criterion \cite{Kass1995}, to approximately \myOrderqty{1d2}{\s} for nested sampling algorithms \cite{Skilling2006a,Feroz2009a,Speagle2020a,Buchner2021c,Ashton2022a}, with Laplace approximations \cite{Nelson2020} and Markov chain Monte Carlo (MCMC)-based estimators \cite{Perrakis2014,Metodiev2024,Llorente2023} spanning intermediate computational regimes. Since all candidate mixtures within a given iteration order $k$ are statistically independent, their evidence evaluations can be performed in parallel. Consequently, assuming sufficient computational resources are available, the wall-clock execution time of BEAP is \mymatht{D \times \Delta t_{\mathcal{Z}}}.

\subsection{Experimental validation}
\label{sub:ExpValMethod}

\noindent To experimentally validate the BEAP algorithm, a series of neutron recoil spectroscopy measurements were performed under laboratory conditions using two neutron sources with distinct spectral characteristics: a {\myCfiso} spontaneous fission neutron source and a deuterium--deuterium (DD) neutron generator (DD P383 Thermo Fischer Scientific), hereafter referred to as {\myCf} and {\myDD}, respectively. The measurements were designed to evaluate the capability of the proposed framework to correctly identify and disentangle neutron source contributions in both isolated and mixed-field configurations.

All neutron fields were measured using the same spectrometer system and acquisition chain employed in our previous work \cite{Breitenmoser2025j} to ensure direct cross-comparability and reproducibility between datasets. The detector system consisted of an array of twelve pulse-shape-discrimination-capable organic-glass scintillator bars with a combined active volume of \qty{21.6}{\cubic\cm}. Data were acquired in full-waveform acquisition mode with a total acquisition time of \qty{\sim2d3}{\s} per experiment. Detailed descriptions of the employed calibration and data reduction pipelines, spectrometer specifications, and background characterization are provided in Ref.~\cite{Breitenmoser2025j}.

\begin{figure}
\centering
\includegraphics[width=0.5\linewidth]{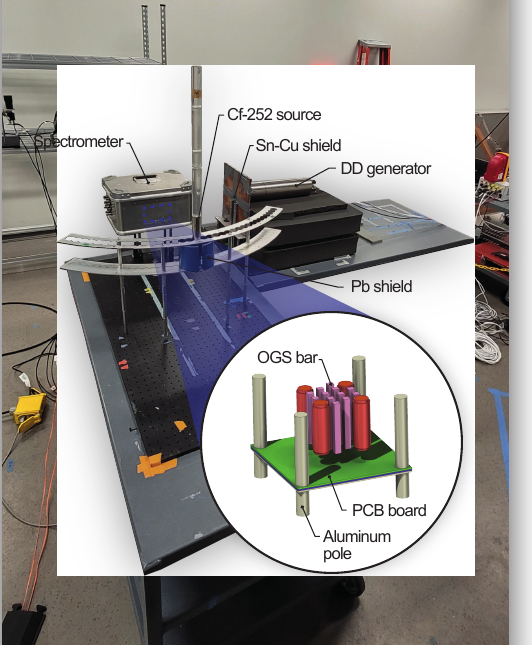}
\caption{Experimental setup used for the validation measurements showing the {\myCf} spontaneous fission source and the {\myDD} neutron generator together with their respective photon shielding, and the organic-glass scintillator (OGS)-based neutron spectrometer.}
\label{fig:ExpSetup}
\end{figure}

Three independent experiments were conducted corresponding to the complete power set of the neutron source mixture considered, namely the single-source configurations \{{\myCf}\} and \{{\myDD}\}, as well as the combined-source configuration \{{\myCf} , {\myDD}\}. Across all measurements, the spatial positions of the individual sources relative to the detector remained fixed in order to ensure geometrical consistency and preserve identical source-to-detector response characteristics between the isolated and mixed-field experiments. A schematic visualization of the experimental configuration for the combined-source measurement containing both {\myCf} and {\myDD} is shown in \cref{fig:ExpSetup}. The single-source experiments were conducted using identical detector and acquisition settings while removing the respective complementary source from the setup.

The {\myCf} source had a neutron emission rate of approximately \qty{1.5(1)E6}{\per\s} and was positioned at a distance of \qty{58}{cm} from the active volume of the spectrometer. To reduce the gamma-ray background inherent to the {\myCf} spontaneous fission source, the source was enclosed in a \qty{6.8}{\mm} lead sleeve. The {\myDD} neutron generator target plane was likewise positioned at a distance of \qty{58}{cm} from the detector active volume. In order to reduce the X-ray background produced during generator operation, the generator head was shielded using a layered \qty{4}{\mm}/\qty{4}{\mm} copper--tin shield assembly. The generator was operated in continuous-wave mode at an acceleration voltage of \qty{80}{\kV} and a beam current of \qty{60}{\micro\A}, corresponding to an approximate neutron emission rate of \myOrderqty{1d6}{\per\s}.

\subsection{Monte Carlo based template generation}
\label{sub:MCmodel}

\begin{figure}
\centering
\includegraphics[width=0.5\linewidth]{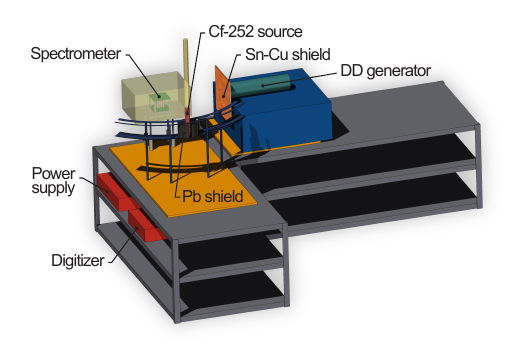}
\caption{Monte Carlo mass model of the experimental setup used for spectral template generation. The model includes the benchmarked organic-glass scintillator spectrometer together with the dominant surrounding experimental components and laboratory structures relevant for neutron transport and scattering.}
\label{fig:SimSetup}
\end{figure}

\noindent  Motivated by analogous physics-based approaches in gamma-ray spectroscopy \cite{Prettyman2006a,Prettyman2011a,Peplowski2016c,Breitenmoser2025m,Breitenmoser2022}, the spectral templates introduced in \cref{sub:BayesMethod} were generated using high-fidelity Monte Carlo neutron transport simulations. The objective of the simulation campaign was to produce physically consistent detector response templates for a representative set of neutron source classes while accurately accounting for detector geometry, surrounding laboratory structures, and neutron scattering effects.

The simulations were performed using the multi-purpose Monte Carlo radiation transport code \texttt{MCNPX-PoliMi}, Version \texttt{2.0} \cite{Pozzi2012}, which provides detailed elastic and inelastic event-by-event neutron interaction tracking and detector-response modeling capabilities. The simulated source library consisted of spontaneous fission sources ({\myCf} \& {\myPu}), \mymatht{(\alpha,\text{n})} neutron sources ({\myPuBe} \& {\myAmBe}), as well as fusion-based deuterium-deuterium ({\myDD}) and deuterium-tritium ({\myDT}) neutron generator sources. The spontaneous fission and \mymatht{(\alpha,\text{n})} neutron sources were modeled as isotropic point emitters positioned at the same source location used for the experimental {\myCf} measurements described in \cref{sub:ExpValMethod}. In contrast, following the modeling protocol by \citet{Lang2018}, the fusion neutron sources were modeled as anisotropic neutron emitters with a \qty{2}{\percent} Doppler broadened emission spectrum rather than isotropic point sources, in order to capture the angular dependence produced by ion acceleration and fusion reactions within the generator target plane \cite{Liskien1973}. For all considered neutron sources, the Monte Carlo simulations employed fully coupled neutron--gamma-ray transport and accounted for all relevant interaction mechanisms in the detector medium, including elastic scattering and the inelastic \mymatht{(\text{n},\text{n}')}, \mymatht{(\text{n},\alpha)}, and \mymatht{(\text{n},\text{n}' \alpha)} reaction channels.

To obtain realistic detector-response predictions, the source descriptions were incorporated into a high-fidelity Monte Carlo mass model of the experimental configuration, shown in \cref{fig:SimSetup}. In addition to the previously benchmarked spectrometer model \cite{Lopez2025}, the simulation geometry included the major structural and experimental components present during the measurements, including high-fidelity neutron generator tube model, optical breadboards, source holders, laboratory furniture, power supplies, digitizer electronics, and the surrounding laboratory room. These structures were included to capture room-return neutrons and secondary scattering contributions that influence the measured recoil spectra. To maintain computational tractability, the individual components were represented as homogeneous media with elemental compositions derived from \citet{McConn2011a}. For each modeled structure, care was taken to preserve the effective material opacity and bulk mass density.

All simulations were executed on the Great Lakes High-Performance-Computing (HPC) cluster at the University of Michigan, comprising more than \num{2d4} computational cores operating at a minimum clock frequency of \qty{2.4}{\GHz}. For each source template, a total of \num{4d10} primary particle histories were simulated to ensure sufficiently low statistical uncertainty (\qty{<5}{\percent}) in the generated detector response spectra. Detailed specifications of the employed calibration and data reduction pipelines are available in our prior work \cite{Breitenmoser2025j,Lopez2025}.

\subsection{Data augmentation}
\label{sub:DataAugmentation}

\noindent  To enable comprehensive testing and statistical validation of the proposed source identification framework, an extensive synthetic dataset of neutron recoil spectra was generated using the Monte Carlo-derived source templates described in \cref{sub:MCmodel}. The objective of the data augmentation procedure was to systematically evaluate the performance and robustness of the BEAP algorithm across a broad range of source mixture configurations and stochastic realizations beyond the experimentally accessible measurements.

Specifically, synthetic recoil spectra were generated for an exhaustive set of single-source, two-source, and three-source mixtures constructed from the predefined neutron source library introduced in \cref{sub:MCmodel}. For each source combination, the expected detector response was computed according to the forward model formulation introduced in \cref{eq:forwardmodelBEAP}. To generalize the comparison across detector configurations and measurement setups, the simulated spectra were scaled by the number of detected neutron events rather than by a measurement live time. The mean recoil spectrum was then obtained as the linear superposition of the source-specific detector-response templates, weighted by the prescribed neutron emission strengths, which ranged from \qty{1d4}{\per\s} to \qty{1d6}{\per\s}. The total number of detected neutron events was varied from \num{1} to \num{1e9}, corresponding to measurement live times between \qty{\sim1d-2}{\s} and \qty{\sim1d7}{\s} for the adopted experimental setup.

To realistically reproduce the statistical dispersion observed in experimental neutron recoil spectroscopy measurements, the computed mean detector response was subsequently propagated through the probabilistic forward model introduced in \cref{sub:BayesMethod}. The resulting synthetic datasets therefore incorporate both the physics-based detector response obtained from Monte Carlo transport simulations and realistic statistical fluctuations representative of experimentally measured neutron recoil spectra. Each generated spectrum was stored together with its associated ground-truth source composition and source strengths, enabling quantitative benchmarking of source identification accuracy, uncertainty quantification, and Bayesian model selection performance of the proposed BEAP framework.

\section{Results}
\label{sec:Results}

\subsection{Experimental Validation}
\label{sub:ResultsVal}

\begin{figure}
\centering
\includegraphics[width=0.95\linewidth]{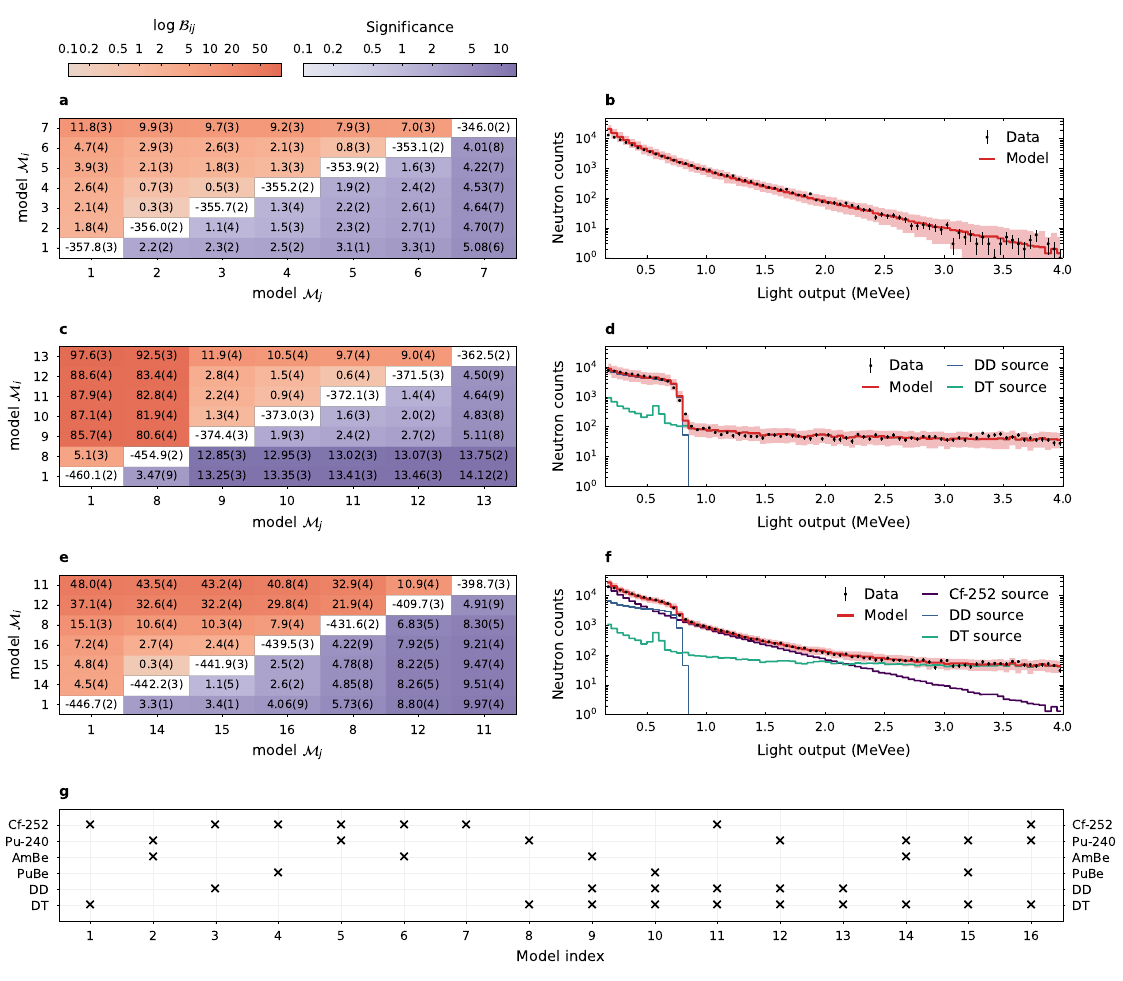}
\caption{Bayesian inference results for the neutron recoil spectroscopy experiments performed using single- and mixed-source configurations. (a--b) Single-source {\myCf} experiment. (c--d) Single-source {\myDD} experiment. (e--f) Two-source {\myCf} \& {\myDD} experiment. Panels (a), (c), and (e) summarize the Bayesian model comparison results for the seven highest-ranking source mixture models evaluated by the BEAP algorithm. Shown are the logarithmic evidence values \mymatht{\log\mathcal{Z}_{i=j}} for each tested model \mymatht{\mathcal{M}_{i=j}} (anti-diagonal entries), the corresponding logarithmic Bayes factors \mymatht{\log\mathcal{B}_{ij}} (entries above the anti-diagonal; see \cref{sub:BayesMethod}), and lower bounds on the statistical significance expressed as Gaussian-equivalent significance levels in units of standard deviations (entries below the anti-diagonal) \cite{Sellke2001,trotta2008a}. Reported uncertainties are given using least-significant-figure notation. Panels (b), (d), and (f) show the posterior predictive distributions for the inferred source mixtures together with the experimentally measured energy spectra. Experimental uncertainties are shown as 1~standard errors. The posterior predictive distributions are visualized using the posterior median prediction together with the corresponding \qty{99}{\percent} central posterior predictive interval (shaded region). In addition, the individual spectral signatures of all sources included in the respective forward model are shown after scaling by the inferred posterior median source strengths and the experimental live time. Panel (g) specifies the tested source mixtures reported in the panels (a), (c), and (e).}
\label{fig:ExpResults}
\end{figure}

\noindent The Bayesian evidence results for the validation experiments described in \cref{sub:ExpValMethod} are summarized in \cref{fig:ExpResults}. For each experiment, the seven highest-ranking source mixture models evaluated by the BEAP algorithm are reported. Panels (a), (c), and (e) show the Bayesian model comparison results for the single-source {\myCf}, single-source {\myDD}, and mixed-source {\myCf} \& {\myDD} experiments, respectively. In each panel, the tested source mixture models are ordered according to decreasing Bayesian evidence. The anti-diagonal entries report the logarithmic evidence values, \mymatht{\log\mathcal{Z}}, for the corresponding models. Entries above the anti-diagonal give the pairwise logarithmic Bayes factors, \mymatht{\log\mathcal{B}_{ij}}, comparing the evidence support between competing models $i$ and $j$, while entries below the anti-diagonal provide lower bounds on the corresponding Gaussian-equivalent statistical significance. Thus, these panels jointly summarize both the absolute evidence ranking of the candidate mixtures and the relative statistical separation between them. For all three validation experiments, the source mixture associated with the experimental configuration was recovered as the highest-evidence model. In each case, the preferred model was favored over the second-highest-ranking alternative with decisive statistical support on Jeffreys' scale (\mymatht{\log\mathcal{B}>\num{9.0}}), corresponding to a Gaussian-equivalent statistical significance of at least \mymatht{4\upsigma} \cite{trotta2008a,Jeffreys1948}. This demonstrates that the BEAP framework correctly identifies the dominant neutron source contributions for both isolated and mixed-field measurements.

Beyond identifying the primary neutron sources, the evidence ranking also resolves weaker secondary components. In the measurements involving the {\myDD} neutron generator, the posterior predictive spectra in panels (d) and (f) of \cref{fig:ExpResults} reveal a reproducible high-energy recoil plateau. This feature is characteristic of the \qty{\sim14.1}{\MeV} neutron component produced by {\myDT} fusion reactions and is consistent with the known high-energy neutron background of nominal {\myDD} generators, which can arise from tritium production, retention, or contamination within the generator target assembly \cite{Lang2018,Cecil1986}. Alternative active-source explanations are disfavored because none of the other source templates considered in the library exhibit comparable high-energy spectral structure. An explanation in terms of ambient or instrumental background is likewise unlikely, given the measured signal-to-background ratios of \num{>1d4} across the evaluated spectral range \cite{Breitenmoser2025j}. Consistent with this spectral signature, BEAP favors source mixtures containing a minor {\myDT} contribution, with an inferred {\myDT}-to-total neutron emission-rate ratio of \num{0.25(0.04:0.03)}, reported as posterior median with corresponding \qty{99}{\percent} credible interval. The statistically significant recovery of the DT contribution demonstrates that the proposed framework can identify not only dominant source classes but also weaker contamination signatures with distinguishable spectral structures.

The experimental validation results above were obtained for a limited set of controlled source mixtures with comparable neutron emission rates and a fixed acquisition time of \qty{\sim2d3}{\s} per experiment. While these measurements demonstrate the ability of BEAP to recover both dominant and weak secondary source components under realistic laboratory conditions, they do not by themselves fully characterize the dependence of the method on source-mixture complexity, counting statistics, or relative source strengths. To systematically assess these effects, the following sections use synthetic recoil spectra generated from known source compositions and emission rates. The use of synthetic recoil spectra enables benchmarking against an exact ground truth, both in terms of the active source set and the underlying source strengths, while allowing exhaustive tests over single-, two-, and three-source mixtures and controlled variations in the number of detected neutron events and multi-source emission-rate ratios.

\subsection{Source Mixture Complexity}
\label{sub:ResultsComplex}

\begin{figure}
\centering
\includegraphics[width=14cm]{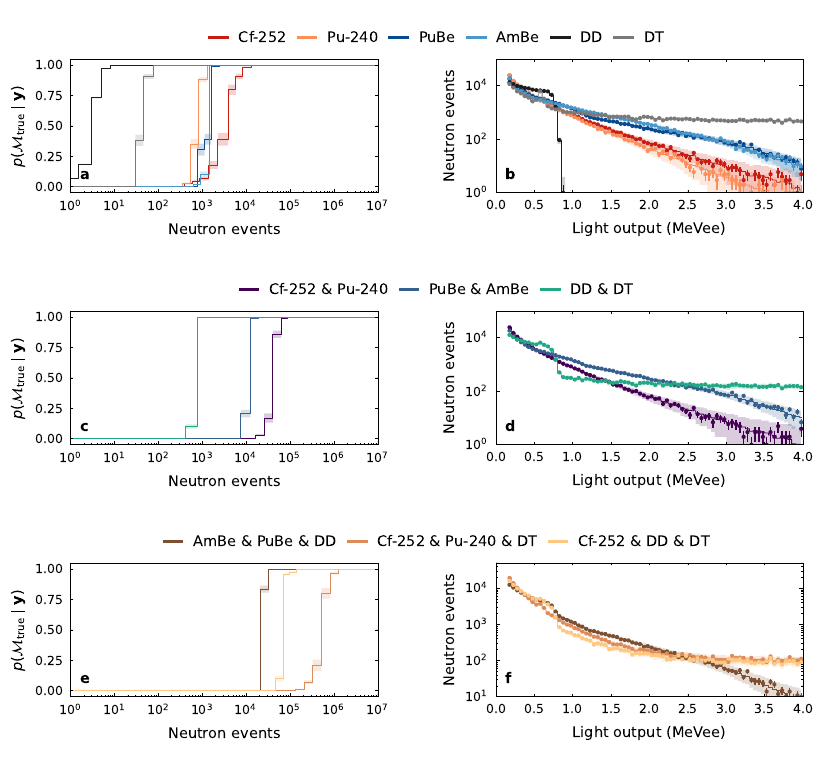}
\caption{Bayesian inference results for synthetic neutron recoil spectroscopy datasets for varying source mixtures and neutron events: (a--b) Single-source mixtures \{{\myCf}\}, \{{\myPu}\}, \{{\myPuBe}\}, \{{\myAmBe}\}, \{{\myDD}\}, and \{{\myDT}\}; (c--d) Two-source mixtures \{{\myCf}, {\myPu}\}, \{{\myPuBe}, {\myAmBe}\}, and \{{\myDD}, {\myDT}\}; (e--f) Three-source mixtures \{{\myAmBe}, {\myPuBe}, {\myDD}\}, \{{\myCf}, {\myPu}, {\myDT}\}, and \{{\myCf}, {\myDD}, {\myDT}\}.  Panels (a), (c), and (e) show the posterior model probability \mymatht{p(\mathcal{M}_\mathrm{true}\mid\myvect{y})} of retrieving the true source model \mymatht{\mathcal{M}_\mathrm{true}} as a function of the total number of detected events in spectrum \mymatht{\myvect{y}}. Uncertainties are indicated as 3-sigma shaded areas. Panels (b), (d), and (f) show posterior predictive distributions for the inferred source mixtures together with the synthetic energy spectra for \num{1d5} neutron events. Synthetic spectrum uncertainties are shown as 1~standard errors. The posterior predictive distributions are visualized using the posterior median prediction together with the corresponding \qty{99}{\percent} central posterior predictive interval (shaded region).}
\label{fig:SynSrc}
\end{figure}

\noindent We first isolate the effect of source-mixture complexity on model-selection performance for synthetic datasets generated with equal neutron source emission rates of \qty{1d6}{\per\s}, comparable to the experimental validation measurements. To assess the scaling of source-identification fidelity with the total number of detected events, \cref{fig:SynSrc} reports the posterior probability \mymatht{p(\mathcal{M}_\mathrm{true}\mid\myvect{y})} of retrieving the true source model \mymatht{\mathcal{M}_\mathrm{true}} conditioned on the synthetic recoil-spectra \mymatht{\myvect{y}} for single-source, two-source, and three-source mixtures \cite{Breitenmoser2025j}.

As expected, the number of detected events required for confident model identification increases with source-mixture complexity. Across the tested cases, confident recovery of the true model requires event counts ranging from \myOrdernum{1d1} for the most distinguishable single-source cases to approximately \myOrdernum{1d6} for the more challenging three-source mixtures. However, the required event count is not determined by mixture dimensionality alone. Within each source-complexity class, we observe a substantial spread, spanning roughly two to three orders of magnitude, in the number of events required for high-confidence identification. We attribute this spread primarily to the degree of spectral similarity between competing source templates. Source mixtures composed of spectrally distinct components can be identified with comparatively few events, whereas mixtures containing sources with similar recoil signatures require substantially higher statistics to overcome model degeneracies. This interpretation is supported by the posterior predictive spectra shown in panels (b), (d), and (f) of \cref{fig:SynSrc}. For example, in the single-source cases, the fusion sources produce more distinctive recoil signatures than the broad-spectrum fission and \mymatht{(\alpha,\mathrm{n})} sources, leading to faster convergence of \mymatht{p(\mathcal{M}_\mathrm{true}\mid\myvect{y})} toward unity. Similar trends are observed for the two-source and three-source mixtures, where source combinations with more overlapping spectral structure require larger detected-event populations for decisive model discrimination.

\begin{figure}
\centering
\includegraphics[width=0.7368\linewidth]{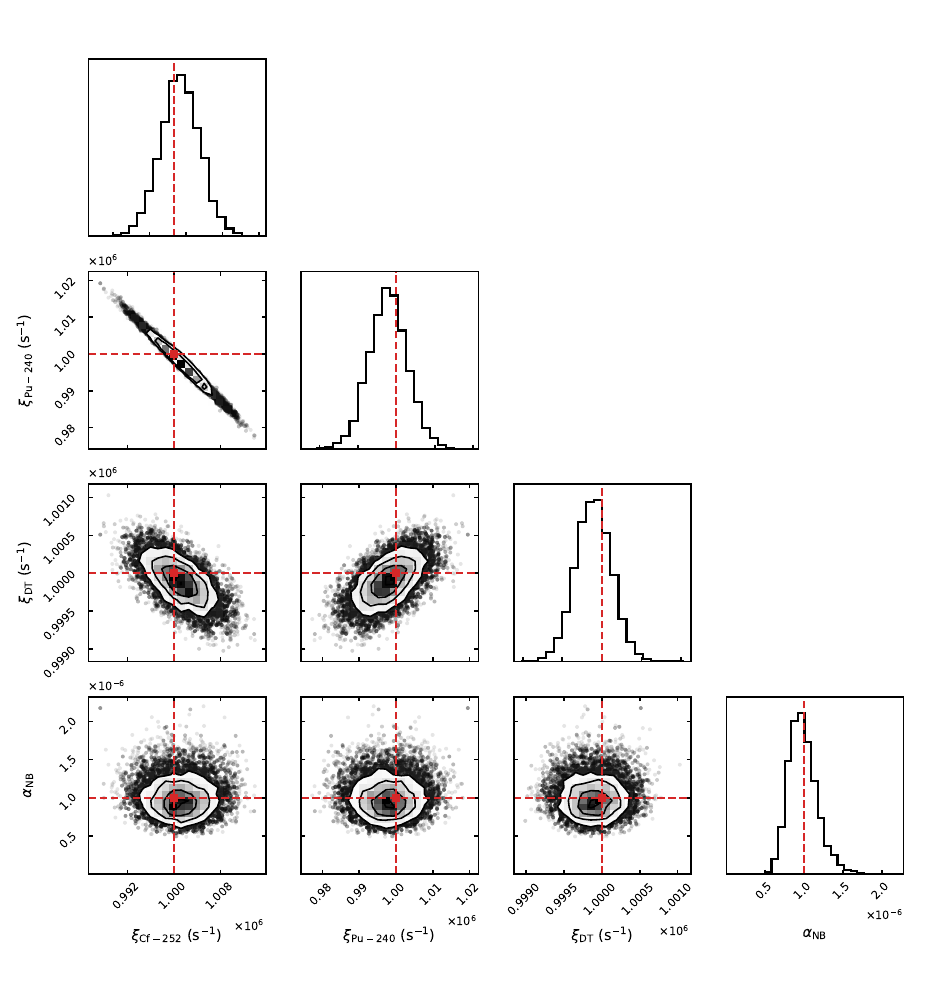}
\caption{Posterior distribution of the \{{\myCf},{\myPu},{\myDT}\} source set for \num{1d5} neutron events. The true neutron emission rates ({\myCfstr},{\myPustr},{\myDTstr}) and the true dispersion parameter ({\mydispersion}) are indicated by the red dashed lines. For the two-dimensional marginal posteriors, we indicate the (\qty{11.8}{\percent}, \qty{39.3}{\percent}, \qty{67.5}{\percent}, \qty{86.4}{\percent}) central credible regions by solid black isolines, corresponding to (0.5, 1, 1.5, 2)-sigma credible regions for a two-dimensional Gaussian distribution \cite{Foreman-Mackey2016}.}
\label{fig:SynCorner}
\end{figure}

It is also worth noting that, when the source--detector geometry is known a priori, the same framework provides accurate posterior estimates of the neutron source emission rates. This condition is important because unknown source positions or source--detector distances would introduce additional normalization degeneracies that prevent direct conversion of detected recoil rates into absolute emission-rate estimates. As an example, \cref{fig:SynCorner} shows the posterior distribution for the challenging three-source mixture \mymatht{\{{\myCf},{\myPu},{\myDT}\}} at \num{1d5} detected neutron events. Despite the increased model complexity and spectral overlap, the true source emission rates and dispersion parameter are recovered without appreciable bias, with sub-percent-level posterior constraints. Complete posterior distribution results for all synthetic datasets are provided in the Supplementary Materials.

\subsection{Neutron Emission Rate Dependence}
\label{sub:ResultsStrength}

\begin{figure}
\centering
\includegraphics[width=14cm]{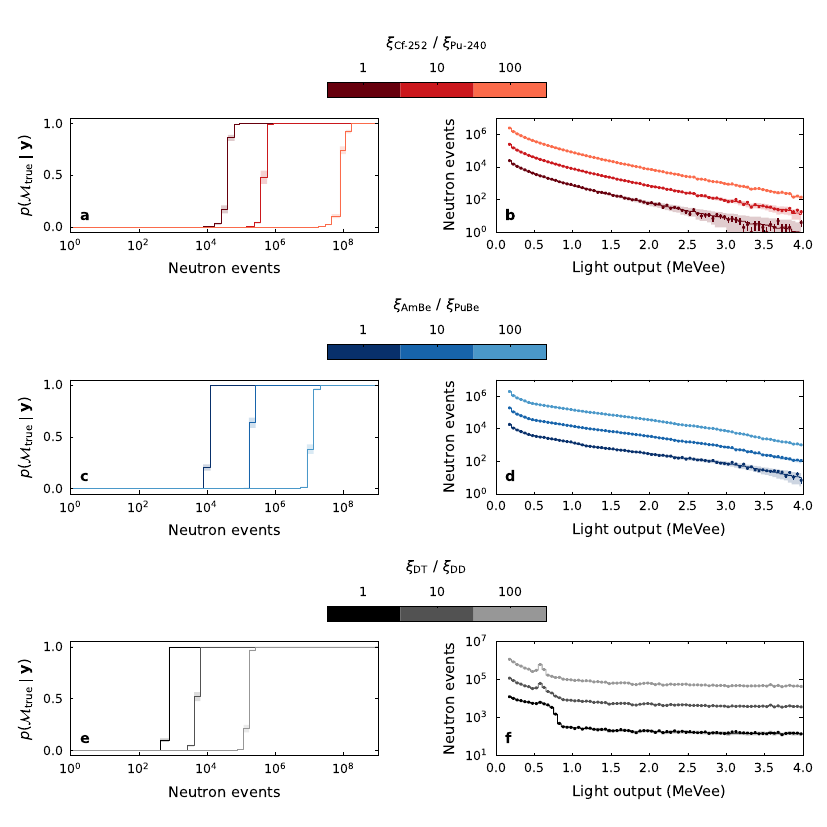}
\caption{Bayesian inference results for synthetic neutron recoil spectroscopy datasets for varying neutron emission rate ratios \mymathtv{\xi_i/\xi_j} for to-source mixtures \{$i$, $j$\} and varying neutron events: (a--b) Two-source mixture \{{\myCf}, {\myPu}\}; (c--d) Two-source mixture \{{\myPuBe}, {\myAmBe}\}; (e--f) Two-source mixture \{{\myDD}, {\myDT}\}. Panels (a), (c), and (e) show the posterior model probability \mymatht{p(\mathcal{M}_\mathrm{true}\mid\myvect{y})} of retrieving the true source model \mymatht{\mathcal{M}_\mathrm{true}} as a function of the total number of detected events in spectrum \mymatht{\myvect{y}}. Uncertainties are indicated as 3-sigma shaded areas. Panels (b), (d), and (f) show posterior predictive distributions for the inferred source mixtures together with the synthetic energy spectra. For better interpretability, the energy spectra are indicated for an equivalent total number of neutron events of \mymathtv{\num{1d5}\times\xi_i/\xi_j}, with \mymathtv{\xi_i/\xi_j} being the neutron emission ratio between source $i$ and source $j$. Synthetic spectrum uncertainties are shown as 1~standard errors. The posterior predictive distributions are visualized using the posterior median prediction together with the corresponding \qty{99}{\percent} central posterior predictive interval (shaded region).}
\label{fig:SynStrength}
\end{figure}

\noindent We next relax the assumption of comparable neutron source emission rates and examine how source-identification fidelity depends on the relative strength of the contributing sources. Specifically, we generated synthetic recoil spectra for representative two-source mixtures with neutron emission-rate ratios varied from unity to \num{1d2}. For each synthetic dataset, BEAP performed Bayesian model selection over the complete candidate library of single-, two-, and three-source mixtures, such that the true generating source pair had to be recovered from the full set of competing source hypotheses rather than being assumed a priori. This analysis probes the ability of BEAP to recover weak source contributions in the presence of an increasingly dominant co-emitted neutron field, representative of an unknown neutron background or interfering source. The resulting posterior model probabilities are shown in \cref{fig:SynStrength}. Panels (a), (c), and (e) report the posterior probability \mymatht{p(\mathcal{M}_\mathrm{true}\mid\myvect{y})} of retrieving the true source model \mymatht{\mathcal{M}_\mathrm{true}} conditioned on the synthetic recoil-spectrum \mymatht{\myvect{y}} as a function of the total number of detected events. The tested source pairs correspond to the two-source mixtures \mymatht{\{{\myCf},{\myPu}\}}, \mymatht{\{{\myPuBe},{\myAmBe}\}}, and \mymatht{\{{\myDD},{\myDT}\}}.

As expected, increasing the imbalance between the neutron source emission rates leads to a systematic increase in the number of detected events required for high-confidence identification of the true source mixture. This behavior arises because the spectral contribution of the weaker source becomes progressively smaller relative to the dominant component, requiring higher counting statistics for its recoil signature to be distinguished from statistical fluctuations and from template degeneracies. The effect is particularly pronounced for source pairs with similar spectral structure, consistent with the trends observed in \cref{sub:ResultsComplex}. This interpretation is further supported by the posterior predictive spectra shown in panels (b), (d), and (f) of \cref{fig:SynStrength}. As the emission-rate ratio increases, the recoil spectrum becomes increasingly dominated by the stronger source, while the weaker component contributes only a subtle distortion to the total spectral shape. Consequently, source pairs with more distinct recoil signatures remain identifiable at lower event counts, whereas spectrally similar pairs require substantially larger detected-event populations for decisive model discrimination. These results demonstrate that the proposed Bayesian framework can recover multi-source mixtures even under strongly non-uniform source contributions, provided that sufficient counting statistics are available.

\section{Conclusion}
\label{sec:Conclusion}

\noindent In this work, we introduced a Bayesian Evidence Adaptive Pursuit framework for neutron source identification from recoil spectroscopy measurements. Building on our previous Bayesian full-spectrum inference approach \cite{Breitenmoser2025j}, the present study addresses two key limitations: the reliance on experimentally trained data-driven forward models and the exponential cost of exhaustive source-mixture enumeration. By replacing the learned forward model with Monte Carlo-generated physics-based response templates and introducing an adaptive evidence-guided search strategy, the proposed framework enables statistically rigorous source identification while improving scalability to larger candidate source libraries and more complex multi-source mixtures.

Experimental validation with {\myCf} and {\myDD} neutron sources showed that the BEAP identification algorithm recovered the correct dominant source configuration with decisive statistical significance (\mymatht{>\!4\upsigma}) in all tested cases. The framework also resolved a secondary {\myDT} contribution in the nominal {\myDD} generator data, consistent with the observed high-energy recoil plateau from \qty{\sim14.1}{\MeV} neutrons, demonstrating sensitivity to both dominant source classes and weaker contamination signatures. Synthetic studies showed that the detected-event requirements increase with source-mixture complexity, spectral similarity, and emission-rate imbalance. For the source library and experimental setup considered here, confident identification under comparable neutron emission rates required approximately \myOrdernum{1d4}, \myOrdernum{1d5}, and \myOrdernum{1d6} detected events for the most challenging single-, two-, and three-source mixtures, respectively.

Translating these detector-agnostic event-count requirements into acquisition times depends directly on the detection efficiency and source--detector geometry. For the experimental configuration used here---a \qty{21.6}{\cubic\cm} organic-glass scintillator spectrometer operated at source--detector separations of \qty{\sim0.5}{\m} with neutron emission rates of \myOrderqty{1d6}{\per\s}---the measured neutron count rates were \qty{\sim50}{\per\s}. The found event-count thresholds therefore correspond to approximate acquisition times of \qty{\sim2d2}{\s}, \qty{\sim2d3}{\s}, and \qty{\sim2d4}{\s} for the most challenging single-, two-, and three-source mixtures considered, respectively. These values are representative of the relatively low-volume validation spectrometer used in this study rather than intrinsic limits of the proposed method. Higher-efficiency detector systems with larger active volumes (\myOrderqty{1d3}{\cubic\cm}) could increase the effective count rate by \myOrdernum{1d2}, reducing comparable identification tasks, even for challenging three-source mixtures, to minute-scale acquisition times.

These performance estimates highlight the practical potential of the proposed framework, while also identifying the main challenges for deployment in less controlled environments. Here, we restricted the analysis to a representative set of source classes, known source--detector geometries, and laboratory configurations with limited shielding and environmental variability. In operational measurements, unknown source position, source encapsulation, intervening shielding, neutron moderation, anisotropic emission, and room-return scattering can modify the measured recoil spectrum and introduce degeneracies between source identity, source strength, and neutron-field transport effects. Validation of the core Bayesian pipeline under mild variations in source position and source shielding in our previous work demonstrated robust source identification performance with decisive statistical evidence, indicating limited impact for these perturbations \cite{Breitenmoser2025j}. More substantial deviations will require extending the Monte Carlo-derived template library to span these experimental and environmental degrees of freedom, enabling their treatment within the Bayesian forward model.

These performance estimates highlight the practical potential of the proposed framework while also identifying the main challenges for deployment in less controlled environments. Here, we restricted the analysis to a representative set of source classes, known source--detector geometries, and laboratory configurations with limited shielding and environmental variability. In operational measurements, unknown source position, source encapsulation, intervening shielding, neutron moderation, anisotropic emission, and room-return scattering can all modify the measured recoil spectrum and introduce degeneracies between source identity, source strength, and neutron-field transport effects. Validation of the core Bayesian pipeline under mild modifications of source positions (\qty{35}{\degree} difference in azimuth) and source casing (additional \qty{6.8}{\mm} lead casing) performed in our previous work demonstrated robust source identification with decisive statistical evidence, indicating limited sensitivity to these perturbations \cite{Breitenmoser2025j}. However, more substantial deviations from the nominal geometry require extending the Monte Carlo-derived template library to span these additional degrees of freedom. Future work should therefore expand the Monte Carlo-derived template library into a higher-dimensional physics-informed response model that spans these experimental and environmental degrees of freedom. Incorporating this variability directly into the Bayesian forward model would enable joint inference of the active source mixture, emission rates, and physical modulation processes affecting the neutron field before detection. Quantifying template-model uncertainty and propagating Monte Carlo statistical and systematic errors into the evidence calculation will also be important for robust field deployment. These developments will support applications in safeguards, nonproliferation, nuclear forensics, radiochemistry, emergency response, and other scenarios requiring quantitative neutron source attribution.

\clearpage

\printcredits

\section*{Acknowledgments}

\noindent This work was supported in part by the Consortium for Monitoring, Technology, and Verification under U.S. Department of Energy National Nuclear Security Administration Award No. DE-NA0003920. Computational resources and services were provided in part by Advanced Research Computing at the University of Michigan, Ann Arbor (SCR\_027337). 

Sandia National Laboratories is a multimission laboratory managed and operated by National Technology and Engineering Solutions of Sandia LLC, a wholly owned subsidiary of Honeywell International Inc. for the U.S. Department of Energy’s National Nuclear Security Administration under contract DE-NA0003525.

\section*{Data and code availability}

\noindent The \texttt{MCNPX-PoliMi} code \citep{Pozzi2012} used for the Monte Carlo simulations is available through the Radiation Safety Information Computational Center (RSICC): \url{https://rsicc.ornl.gov} (Computer Code Number: C00791). The Postprocessing pipeline \texttt{pyMPPost} adopted for this work is publicly available at GitLab: \url{https://gitlab.eecs.umich.edu/umich-dnng/pymppost}. All experimental data presented in this study have been deposited in the Zenodo repository under accession code \url{https://doi.org/10.5281/zenodo.21516446} \citep{Breitenmoser2026c}.

\bibliographystyle{elsarticle-num-names}

\begin{thebibliography}{48}
\expandafter\ifx\csname natexlab\endcsname\relax\def\natexlab#1{#1}\fi
\providecommand{\url}[1]{\texttt{#1}}
\providecommand{\href}[2]{#2}
\providecommand{\path}[1]{#1}
\providecommand{\DOIprefix}{doi:}
\providecommand{\ArXivprefix}{arXiv:}
\providecommand{\URLprefix}{URL: }
\providecommand{\Pubmedprefix}{pmid:}
\providecommand{\doi}[1]{\href{http://dx.doi.org/#1}{\path{#1}}}
\providecommand{\Pubmed}[1]{\href{pmid:#1}{\path{#1}}}
\providecommand{\bibinfo}[2]{#2}
\ifx\xfnm\relax \def\xfnm[#1]{\unskip,\space#1}\fi
\bibitem[{Al~Hamrashdi et~al.(2019)Al~Hamrashdi, Monk, and Cheneler}]{alhamrashdiPassiveGammaRayNeutron2019}
\bibinfo{author}{H.~Al~Hamrashdi}, \bibinfo{author}{S.~D. Monk}, \bibinfo{author}{D.~Cheneler},
\newblock \bibinfo{title}{Passive {{Gamma-Ray}} and {{Neutron Imaging Systems}} for {{National Security}} and {{Nuclear Non-Proliferation}} in {{Controlled}} and {{Uncontrolled Detection Areas}}: {{Review}} of {{Past}} and {{Current Status}}},
\newblock \bibinfo{journal}{Sensors} \bibinfo{volume}{19} (\bibinfo{year}{2019}) \bibinfo{pages}{2638}. \DOIprefix\doi{10.3390/s19112638}.
\bibitem[{Breitenmoser et~al.(2026)Breitenmoser, Lopez, Clarke, and Pozzi}]{Breitenmoser2025j}
\bibinfo{author}{D.~Breitenmoser}, \bibinfo{author}{R.~Lopez}, \bibinfo{author}{S.~D. Clarke}, \bibinfo{author}{S.~A. Pozzi},
\newblock \bibinfo{title}{Identifying neutron sources using recoil and time-of-flight spectroscopy},
\newblock \bibinfo{journal}{Phys. Rev. Appl.} \bibinfo{volume}{25} (\bibinfo{year}{2026}) \bibinfo{pages}{064013}. \DOIprefix\doi{10.1103/v6j6-f4rx}. \href{http://arxiv.org/abs/2512.10044}{{\tt arXiv:2512.10044}}.
\bibitem[{Wallenius et~al.(2007)Wallenius, L{\"u}tzenkirchen, Mayer, Ray, {de las Heras}, Betti, Cromboom, Hild, Lynch, Nicholl, Ottmar, Rasmussen, Schubert, Tamborini, Thiele, Wagner, Walker, and Zuleger}]{Wallenius2007}
\bibinfo{author}{M.~Wallenius}, \bibinfo{author}{K.~L{\"u}tzenkirchen}, \bibinfo{author}{K.~Mayer}, \bibinfo{author}{I.~Ray}, \bibinfo{author}{L.~A. {de las Heras}}, \bibinfo{author}{M.~Betti}, \bibinfo{author}{O.~Cromboom}, \bibinfo{author}{M.~Hild}, \bibinfo{author}{B.~Lynch}, \bibinfo{author}{A.~Nicholl}, \bibinfo{author}{H.~Ottmar}, \bibinfo{author}{G.~Rasmussen}, \bibinfo{author}{A.~Schubert}, \bibinfo{author}{G.~Tamborini}, \bibinfo{author}{H.~Thiele}, \bibinfo{author}{W.~Wagner}, \bibinfo{author}{C.~Walker}, \bibinfo{author}{E.~Zuleger},
\newblock \bibinfo{title}{Nuclear forensic investigations with a focus on plutonium},
\newblock \bibinfo{journal}{Journal of Alloys and Compounds} \bibinfo{volume}{444--445} (\bibinfo{year}{2007}) \bibinfo{pages}{57--62}. \DOIprefix\doi{10.1016/j.jallcom.2006.10.161}.
\bibitem[{Mayer et~al.(2013)Mayer, Wallenius, and Varga}]{Mayer2013}
\bibinfo{author}{K.~Mayer}, \bibinfo{author}{M.~Wallenius}, \bibinfo{author}{Z.~Varga},
\newblock \bibinfo{title}{Nuclear {{Forensic Science}}: {{Correlating Measurable Material Parameters}} to the {{History}} of {{Nuclear Material}}},
\newblock \bibinfo{journal}{Chem. Rev.} \bibinfo{volume}{113} (\bibinfo{year}{2013}) \bibinfo{pages}{884--900}. \DOIprefix\doi{10.1021/cr300273f}.
\bibitem[{Meng et~al.(2018)Meng, Jianyu, Jun, and Rui}]{Meng2018}
\bibinfo{author}{H.~Meng}, \bibinfo{author}{Z.~Jianyu}, \bibinfo{author}{W.~Jun}, \bibinfo{author}{L.~Rui},
\newblock \bibinfo{title}{A {{Passive Method}} for the {{Detection}} of {{Explosives}} and {{Weapons-Grade Plutonium}} in {{Nuclear Warheads}}},
\newblock \bibinfo{journal}{Sci. Glob. Secur.} \bibinfo{volume}{26} (\bibinfo{year}{2018}) \bibinfo{pages}{57--69}. \DOIprefix\doi{10.1080/08929882.2018.1517431}.
\bibitem[{Lepowsky et~al.(2021)Lepowsky, Jeon, and Glaser}]{Lepowsky2021}
\bibinfo{author}{E.~Lepowsky}, \bibinfo{author}{J.~Jeon}, \bibinfo{author}{A.~Glaser},
\newblock \bibinfo{title}{Confirming the absence of nuclear warheads via passive gamma-ray measurements},
\newblock \bibinfo{journal}{Nucl. Instrum. Methods Phys. Res. A} \bibinfo{volume}{990} (\bibinfo{year}{2021}) \bibinfo{pages}{164983}. \DOIprefix\doi{10.1016/j.nima.2020.164983}.
\bibitem[{LaMont et~al.(1998)LaMont, Glover, and Filby}]{LaMont1998}
\bibinfo{author}{S.~P. LaMont}, \bibinfo{author}{S.~E. Glover}, \bibinfo{author}{R.~H. Filby},
\newblock \bibinfo{title}{Determination of plutonium-240/239 ratios in low activity samples using high resolution alpha-spectrometry},
\newblock \bibinfo{journal}{J Radioanal Nucl Chem} \bibinfo{volume}{234} (\bibinfo{year}{1998}) \bibinfo{pages}{195--199}. \DOIprefix\doi{10.1007/BF02389771}.
\bibitem[{Trotta(2008)}]{trotta2008a}
\bibinfo{author}{R.~Trotta},
\newblock \bibinfo{title}{Bayes in the sky: {{Bayesian}} inference and model selection in cosmology},
\newblock \bibinfo{journal}{Contemp. Phys.} \bibinfo{volume}{49} (\bibinfo{year}{2008}) \bibinfo{pages}{71--104}. \DOIprefix\doi{10.1080/00107510802066753}.
\bibitem[{{von Toussaint}(2011)}]{vonToussaint2011a}
\bibinfo{author}{U.~{von Toussaint}},
\newblock \bibinfo{title}{Bayesian inference in physics},
\newblock \bibinfo{journal}{Rev. Mod. Phys.} \bibinfo{volume}{83} (\bibinfo{year}{2011}) \bibinfo{pages}{943--999}. \DOIprefix\doi{10.1103/RevModPhys.83.943}.
\bibitem[{Veitch(2015)}]{Veitch2015}
\bibinfo{author}{J.~Veitch},
\newblock \bibinfo{title}{Parameter estimation for compact binaries with ground-based gravitational-wave observations using the {{LALInference}} software library},
\newblock \bibinfo{journal}{Phys. Rev. D} \bibinfo{volume}{91} (\bibinfo{year}{2015}). \DOIprefix\doi{10.1103/PhysRevD.91.042003}.
\bibitem[{Ashton et~al.(2019)Ashton, H{\"u}bner, Lasky, Talbot, Ackley, Biscoveanu, Chu, Divakarla, Easter, Goncharov, Vivanco, Harms, Lower, Meadors, Melchor, Payne, Pitkin, Powell, Sarin, Smith, and Thrane}]{Ashton2019}
\bibinfo{author}{G.~Ashton}, \bibinfo{author}{M.~H{\"u}bner}, \bibinfo{author}{P.~D. Lasky}, \bibinfo{author}{C.~Talbot}, \bibinfo{author}{K.~Ackley}, \bibinfo{author}{S.~Biscoveanu}, \bibinfo{author}{Q.~Chu}, \bibinfo{author}{A.~Divakarla}, \bibinfo{author}{P.~J. Easter}, \bibinfo{author}{B.~Goncharov}, \bibinfo{author}{F.~H. Vivanco}, \bibinfo{author}{J.~Harms}, \bibinfo{author}{M.~E. Lower}, \bibinfo{author}{G.~D. Meadors}, \bibinfo{author}{D.~Melchor}, \bibinfo{author}{E.~Payne}, \bibinfo{author}{M.~D. Pitkin}, \bibinfo{author}{J.~Powell}, \bibinfo{author}{N.~Sarin}, \bibinfo{author}{R.~J.~E. Smith}, \bibinfo{author}{E.~Thrane},
\newblock \bibinfo{title}{Bilby: {{A User-friendly Bayesian Inference Library}} for {{Gravitational-wave Astronomy}}},
\newblock \bibinfo{journal}{ApJS} \bibinfo{volume}{241} (\bibinfo{year}{2019}) \bibinfo{pages}{27}. \DOIprefix\doi{10.3847/1538-4365/ab06fc}.
\bibitem[{Smith et~al.(2020)Smith, Ashton, Vajpeyi, and Talbot}]{Smith2020}
\bibinfo{author}{R.~J.~E. Smith}, \bibinfo{author}{G.~Ashton}, \bibinfo{author}{A.~Vajpeyi}, \bibinfo{author}{C.~Talbot},
\newblock \bibinfo{title}{Massively parallel {{Bayesian}} inference for transient gravitational-wave astronomy},
\newblock \bibinfo{journal}{MNRAS} \bibinfo{volume}{498} (\bibinfo{year}{2020}) \bibinfo{pages}{4492--4502}. \DOIprefix\doi{10.1093/mnras/staa2483}.
\bibitem[{MacDonald and Madhusudhan(2017)}]{MacDonald2017}
\bibinfo{author}{R.~J. MacDonald}, \bibinfo{author}{N.~Madhusudhan},
\newblock \bibinfo{title}{{{HD}} 209458b in new light: Evidence of nitrogen chemistry, patchy clouds and sub-solar water},
\newblock \bibinfo{journal}{MNRAS} \bibinfo{volume}{469} (\bibinfo{year}{2017}) \bibinfo{pages}{1979--1996}. \DOIprefix\doi{10.1093/mnras/stx804}.
\bibitem[{Pinhas et~al.(2018)Pinhas, Rackham, Madhusudhan, and Apai}]{Pinhas2018}
\bibinfo{author}{A.~Pinhas}, \bibinfo{author}{B.~V. Rackham}, \bibinfo{author}{N.~Madhusudhan}, \bibinfo{author}{D.~Apai},
\newblock \bibinfo{title}{Retrieval of planetary and stellar properties in transmission spectroscopy with {{Aura}}},
\newblock \bibinfo{journal}{MNRAS} \bibinfo{volume}{480} (\bibinfo{year}{2018}) \bibinfo{pages}{5314--5331}. \DOIprefix\doi{10.1093/mnras/sty2209}.
\bibitem[{H{\"u}nnefeld et~al.(2022)H{\"u}nnefeld, Abbasi, Ackermann, Adams, Aguilar, Ahlers, Ahrens, Alispach, Alves, Amin, An, Andeen, Anderson, Anton, Arg{\"u}elles, Ashida, Axani, Bai, Balagopal, Barbano, Barwick, Bastian, Basu, Baur, Bay, Beatty, Becker, Becker~Tjus, Bellenghi, BenZvi, Berley, Bernardini, Besson, Binder, Bindig, Blaufuss, Blot, Boddenberg, Bontempo, Borowka, B{\"o}ser, Botner, B{\"o}ttcher, Bourbeau, Bradascio, Braun, Bron, {Brostean-Kaiser}, Browne, Burgman, Burley, Busse, Campana, {Carnie-Bronca}, Chen, Chirkin, Choi, Clark, Clark, Classen, Coleman, Collin, Conrad, Coppin, Correa, Cowen, Cross, Dappen, Dave, De~Clercq, DeLaunay, Dembinski, Deoskar, De~Ridder, Desai, Desiati, {de Vries}, {de Wasseige}, {de With}, DeYoung, Dharani, Diaz, {D{\'i}az-V{\'e}lez}, Dittmer, Dujmovic, Dunkman, DuVernois, Dvorak, Ehrhardt, Eller, Engel, Erpenbeck, Evans, Evenson, Fan, Fazely, Fiedlschuster, Fienberg, Filimonov, Finley, Fischer, Fox, Franckowiak, Friedman, Fritz, F{\"u}rst, Gaisser, Gallagher,
  Ganster, Garcia, Garrappa, Gerhardt, Ghadimi, Glaser, Glauch, Gl{\"u}senkamp, Goldschmidt, Gonzalez, Goswami, Grant, Gr{\'e}goire, Griswold, G{\"u}nd{\"u}z, G{\"u}nther, Haack, Hallgren, Halliday, Halve, Halzen, Ha~Minh, Hanson, Hardin, Harnisch, Haungs, Hauser, Hebecker, Helbing, Henningsen, Hettinger, Hickford, Hignight, Hill, Hill, Hoffman, Hoffmann, Hoinka, {Hokanson-Fasig}, Hoshina, Huang, Huber, Huber, Hultqvist, Hussain, In, Iovine, Ishihara, Jansson, Japaridze, Jeong, Jones, Kang, Kang, Kang, Kappes, Kappesser, Karg, Karl, Karle, Katz, Kauer, Kellermann, Kelley, Kheirandish, Kin, Kintscher, Kiryluk, Klein, Koirala, Kolanoski, Kontrimas, K{\"o}pke, Kopper, Kopper, Koskinen, Koundal, Kovacevich, Kowalski, Kozynets, Kun, Kurahashi, Lad, Lagunas~Gualda, Lanfranchi, Larson, Lauber, Lazar, Lee, Leonard, Leszczy{\'n}ska, Li, Lincetto, Liu, Liubarska, Lohfink, Lozano~Mariscal, Lu, Lucarelli, Ludwig, Luszczak, Lyu, Ma, Madsen, Mahn, Makino, Mancina, Mari{\c s}, Maruyama, Mase, McElroy, McNally, Mead,
  Meagher, Medina, Meier, {Meighen-Berger}, Micallef, Mockler, Montaruli, Moore, Morse, Moulai, Naab, Nagai, Naumann, Necker, Nguy{\^e}n, Niederhausen, Nisa, Nowicki, Nygren, Obertacke~Pollmann, Oehler, Olivas, O'Sullivan, Pandya, Pankova, Park, Parker, Paudel, Paul, {P{\'e}rez de los Heros}, Peters, Peterson, Philippen, Pieloth, Pieper, Pittermann, Pizzuto, Plum, Popovych, Porcelli, Prado~Rodriguez, Price, Pries, Przybylski, Raab, Raissi, Rameez, Rawlins, Rea, Rehman, Reichherzer, Reimann, Renzi, Resconi, Reusch, Rhode, Richman, Riedel, Roberts, Robertson, Roellinghoff, Rongen, Rott, Ruhe, Ryckbosch, Rysewyk~Cantu, Safa, Saffer, Sanchez~Herrera, Sandrock, Sandroos, Santander, Sarkar, Sarkar, Satalecka, Scharf, Schaufel, Schieler, Schindler, Schlunder, Schmidt, Schneider, Schneider, Schr{\"o}der, Schumacher, Schwefer, Sclafani, Seckel, Seunarine, Sharma, Shefali, Silva, Skrzypek, Smithers, Snihur, Soedingrekso, Soldin, Spannfellner, Spiczak, Spiering, Stachurska, Stamatikos, Stanev, Stein, Stettner, Steuer,
  Stezelberger, St{\"u}rwald, Stuttard, Sullivan, Taboada, Tenholt, {Ter-Antonyan}, Tilav, Tischbein, Tollefson, Tomankova, T{\"o}nnis, Toscano, Tosi, Trettin, Tselengidou, Tung, Turcati, Turcotte, Turley, Twagirayezu, Ty, Unland~Elorrieta, {Valtonen-Mattila}, Vandenbroucke, {van Eijndhoven}, Vannerom, {van Santen}, Verpoest, Vraeghe, Walck, Watson, Weaver, Weigel, Weindl, Weiss, Weldert, Wendt, Werthebach, Weyrauch, Whitehorn, Wiebusch, Williams, Wolf, Woschnagg, Wrede, Wulff, Xu, Xu, Yanez, Yoshida, Yu, Yuan, and Zhang}]{Hunnefeld2022a}
\bibinfo{author}{M.~H{\"u}nnefeld}, \bibinfo{author}{R.~Abbasi}, \bibinfo{author}{M.~Ackermann}, \bibinfo{author}{J.~Adams}, \bibinfo{author}{J.~A. Aguilar}, \bibinfo{author}{M.~Ahlers}, \bibinfo{author}{M.~Ahrens}, \bibinfo{author}{C.~Alispach}, \bibinfo{author}{A.~A. Alves}, \bibinfo{author}{N.~M. Amin}, \bibinfo{author}{R.~An}, \bibinfo{author}{K.~Andeen}, \bibinfo{author}{T.~Anderson}, \bibinfo{author}{G.~Anton}, \bibinfo{author}{C.~Arg{\"u}elles}, \bibinfo{author}{Y.~Ashida}, \bibinfo{author}{S.~Axani}, \bibinfo{author}{X.~Bai}, \bibinfo{author}{A.~V. Balagopal}, \bibinfo{author}{A.~Barbano}, \bibinfo{author}{S.~W. Barwick}, \bibinfo{author}{B.~Bastian}, \bibinfo{author}{V.~Basu}, \bibinfo{author}{S.~Baur}, \bibinfo{author}{R.~Bay}, \bibinfo{author}{J.~J. Beatty}, \bibinfo{author}{K.~H. Becker}, \bibinfo{author}{J.~Becker~Tjus}, \bibinfo{author}{C.~Bellenghi}, \bibinfo{author}{S.~BenZvi}, \bibinfo{author}{D.~Berley}, \bibinfo{author}{E.~Bernardini}, \bibinfo{author}{D.~Z. Besson},
  \bibinfo{author}{G.~Binder}, \bibinfo{author}{D.~Bindig}, \bibinfo{author}{E.~Blaufuss}, \bibinfo{author}{S.~Blot}, \bibinfo{author}{M.~Boddenberg}, \bibinfo{author}{F.~Bontempo}, \bibinfo{author}{J.~Borowka}, \bibinfo{author}{S.~B{\"o}ser}, \bibinfo{author}{O.~Botner}, \bibinfo{author}{J.~B{\"o}ttcher}, \bibinfo{author}{E.~Bourbeau}, \bibinfo{author}{F.~Bradascio}, \bibinfo{author}{J.~Braun}, \bibinfo{author}{S.~Bron}, \bibinfo{author}{J.~{Brostean-Kaiser}}, \bibinfo{author}{S.~Browne}, \bibinfo{author}{A.~Burgman}, \bibinfo{author}{R.~T. Burley}, \bibinfo{author}{R.~S. Busse}, \bibinfo{author}{M.~A. Campana}, \bibinfo{author}{E.~G. {Carnie-Bronca}}, \bibinfo{author}{C.~Chen}, \bibinfo{author}{D.~Chirkin}, \bibinfo{author}{K.~Choi}, \bibinfo{author}{B.~A. Clark}, \bibinfo{author}{K.~Clark}, \bibinfo{author}{L.~Classen}, \bibinfo{author}{A.~Coleman}, \bibinfo{author}{G.~H. Collin}, \bibinfo{author}{J.~M. Conrad}, \bibinfo{author}{P.~Coppin}, \bibinfo{author}{P.~Correa}, \bibinfo{author}{D.~F. Cowen},
  \bibinfo{author}{R.~Cross}, \bibinfo{author}{C.~Dappen}, \bibinfo{author}{P.~Dave}, \bibinfo{author}{C.~De~Clercq}, \bibinfo{author}{J.~J. DeLaunay}, \bibinfo{author}{H.~Dembinski}, \bibinfo{author}{K.~Deoskar}, \bibinfo{author}{S.~De~Ridder}, \bibinfo{author}{A.~Desai}, \bibinfo{author}{P.~Desiati}, \bibinfo{author}{K.~D. {de Vries}}, \bibinfo{author}{G.~{de Wasseige}}, \bibinfo{author}{M.~{de With}}, \bibinfo{author}{T.~DeYoung}, \bibinfo{author}{S.~Dharani}, \bibinfo{author}{A.~Diaz}, \bibinfo{author}{J.~C. {D{\'i}az-V{\'e}lez}}, \bibinfo{author}{M.~Dittmer}, \bibinfo{author}{H.~Dujmovic}, \bibinfo{author}{M.~Dunkman}, \bibinfo{author}{M.~A. DuVernois}, \bibinfo{author}{E.~Dvorak}, \bibinfo{author}{T.~Ehrhardt}, \bibinfo{author}{P.~Eller}, \bibinfo{author}{R.~Engel}, \bibinfo{author}{H.~Erpenbeck}, \bibinfo{author}{J.~Evans}, \bibinfo{author}{P.~A. Evenson}, \bibinfo{author}{K.~L. Fan}, \bibinfo{author}{A.~R. Fazely}, \bibinfo{author}{S.~Fiedlschuster}, \bibinfo{author}{A.~T. Fienberg},
  \bibinfo{author}{K.~Filimonov}, \bibinfo{author}{C.~Finley}, \bibinfo{author}{L.~Fischer}, \bibinfo{author}{D.~Fox}, \bibinfo{author}{A.~Franckowiak}, \bibinfo{author}{E.~Friedman}, \bibinfo{author}{A.~Fritz}, \bibinfo{author}{P.~F{\"u}rst}, \bibinfo{author}{T.~K. Gaisser}, \bibinfo{author}{J.~Gallagher}, \bibinfo{author}{E.~Ganster}, \bibinfo{author}{A.~Garcia}, \bibinfo{author}{S.~Garrappa}, \bibinfo{author}{L.~Gerhardt}, \bibinfo{author}{A.~Ghadimi}, \bibinfo{author}{C.~Glaser}, \bibinfo{author}{T.~Glauch}, \bibinfo{author}{T.~Gl{\"u}senkamp}, \bibinfo{author}{A.~Goldschmidt}, \bibinfo{author}{J.~G. Gonzalez}, \bibinfo{author}{S.~Goswami}, \bibinfo{author}{D.~Grant}, \bibinfo{author}{T.~Gr{\'e}goire}, \bibinfo{author}{S.~Griswold}, \bibinfo{author}{M.~G{\"u}nd{\"u}z}, \bibinfo{author}{C.~G{\"u}nther}, \bibinfo{author}{C.~Haack}, \bibinfo{author}{A.~Hallgren}, \bibinfo{author}{R.~Halliday}, \bibinfo{author}{L.~Halve}, \bibinfo{author}{F.~Halzen}, \bibinfo{author}{M.~Ha~Minh}, \bibinfo{author}{K.~Hanson},
  \bibinfo{author}{J.~Hardin}, \bibinfo{author}{A.~A. Harnisch}, \bibinfo{author}{A.~Haungs}, \bibinfo{author}{S.~Hauser}, \bibinfo{author}{D.~Hebecker}, \bibinfo{author}{K.~Helbing}, \bibinfo{author}{F.~Henningsen}, \bibinfo{author}{E.~C. Hettinger}, \bibinfo{author}{S.~Hickford}, \bibinfo{author}{J.~Hignight}, \bibinfo{author}{C.~Hill}, \bibinfo{author}{G.~C. Hill}, \bibinfo{author}{K.~D. Hoffman}, \bibinfo{author}{R.~Hoffmann}, \bibinfo{author}{T.~Hoinka}, \bibinfo{author}{B.~{Hokanson-Fasig}}, \bibinfo{author}{K.~Hoshina}, \bibinfo{author}{F.~Huang}, \bibinfo{author}{M.~Huber}, \bibinfo{author}{T.~Huber}, \bibinfo{author}{K.~Hultqvist}, \bibinfo{author}{R.~Hussain}, \bibinfo{author}{S.~In}, \bibinfo{author}{N.~Iovine}, \bibinfo{author}{A.~Ishihara}, \bibinfo{author}{M.~Jansson}, \bibinfo{author}{G.~S. Japaridze}, \bibinfo{author}{M.~Jeong}, \bibinfo{author}{B.~J. Jones}, \bibinfo{author}{D.~Kang}, \bibinfo{author}{W.~Kang}, \bibinfo{author}{X.~Kang}, \bibinfo{author}{A.~Kappes},
  \bibinfo{author}{D.~Kappesser}, \bibinfo{author}{T.~Karg}, \bibinfo{author}{M.~Karl}, \bibinfo{author}{A.~Karle}, \bibinfo{author}{U.~Katz}, \bibinfo{author}{M.~Kauer}, \bibinfo{author}{M.~Kellermann}, \bibinfo{author}{J.~L. Kelley}, \bibinfo{author}{A.~Kheirandish}, \bibinfo{author}{K.~Kin}, \bibinfo{author}{T.~Kintscher}, \bibinfo{author}{J.~Kiryluk}, \bibinfo{author}{S.~R. Klein}, \bibinfo{author}{R.~Koirala}, \bibinfo{author}{H.~Kolanoski}, \bibinfo{author}{T.~Kontrimas}, \bibinfo{author}{L.~K{\"o}pke}, \bibinfo{author}{C.~Kopper}, \bibinfo{author}{S.~Kopper}, \bibinfo{author}{D.~J. Koskinen}, \bibinfo{author}{P.~Koundal}, \bibinfo{author}{M.~Kovacevich}, \bibinfo{author}{M.~Kowalski}, \bibinfo{author}{T.~Kozynets}, \bibinfo{author}{E.~Kun}, \bibinfo{author}{N.~Kurahashi}, \bibinfo{author}{N.~Lad}, \bibinfo{author}{C.~Lagunas~Gualda}, \bibinfo{author}{J.~L. Lanfranchi}, \bibinfo{author}{M.~J. Larson}, \bibinfo{author}{F.~Lauber}, \bibinfo{author}{J.~P. Lazar}, \bibinfo{author}{J.~W. Lee},
  \bibinfo{author}{K.~Leonard}, \bibinfo{author}{A.~Leszczy{\'n}ska}, \bibinfo{author}{Y.~Li}, \bibinfo{author}{M.~Lincetto}, \bibinfo{author}{Q.~R. Liu}, \bibinfo{author}{M.~Liubarska}, \bibinfo{author}{E.~Lohfink}, \bibinfo{author}{C.~J. Lozano~Mariscal}, \bibinfo{author}{L.~Lu}, \bibinfo{author}{F.~Lucarelli}, \bibinfo{author}{A.~Ludwig}, \bibinfo{author}{W.~Luszczak}, \bibinfo{author}{Y.~Lyu}, \bibinfo{author}{W.~Y. Ma}, \bibinfo{author}{J.~Madsen}, \bibinfo{author}{K.~B. Mahn}, \bibinfo{author}{Y.~Makino}, \bibinfo{author}{S.~Mancina}, \bibinfo{author}{I.~C. Mari{\c s}}, \bibinfo{author}{R.~Maruyama}, \bibinfo{author}{K.~Mase}, \bibinfo{author}{T.~McElroy}, \bibinfo{author}{F.~McNally}, \bibinfo{author}{J.~V. Mead}, \bibinfo{author}{K.~Meagher}, \bibinfo{author}{A.~Medina}, \bibinfo{author}{M.~Meier}, \bibinfo{author}{S.~{Meighen-Berger}}, \bibinfo{author}{J.~Micallef}, \bibinfo{author}{D.~Mockler}, \bibinfo{author}{T.~Montaruli}, \bibinfo{author}{R.~W. Moore}, \bibinfo{author}{R.~Morse},
  \bibinfo{author}{M.~Moulai}, \bibinfo{author}{R.~Naab}, \bibinfo{author}{R.~Nagai}, \bibinfo{author}{U.~Naumann}, \bibinfo{author}{J.~Necker}, \bibinfo{author}{L.~V. Nguy{\^e}n}, \bibinfo{author}{H.~Niederhausen}, \bibinfo{author}{M.~U. Nisa}, \bibinfo{author}{S.~C. Nowicki}, \bibinfo{author}{D.~R. Nygren}, \bibinfo{author}{A.~Obertacke~Pollmann}, \bibinfo{author}{M.~Oehler}, \bibinfo{author}{A.~Olivas}, \bibinfo{author}{E.~O'Sullivan}, \bibinfo{author}{H.~Pandya}, \bibinfo{author}{D.~V. Pankova}, \bibinfo{author}{N.~Park}, \bibinfo{author}{G.~K. Parker}, \bibinfo{author}{E.~N. Paudel}, \bibinfo{author}{L.~Paul}, \bibinfo{author}{C.~{P{\'e}rez de los Heros}}, \bibinfo{author}{L.~Peters}, \bibinfo{author}{J.~Peterson}, \bibinfo{author}{S.~Philippen}, \bibinfo{author}{D.~Pieloth}, \bibinfo{author}{S.~Pieper}, \bibinfo{author}{M.~Pittermann}, \bibinfo{author}{A.~Pizzuto}, \bibinfo{author}{M.~Plum}, \bibinfo{author}{Y.~Popovych}, \bibinfo{author}{A.~Porcelli}, \bibinfo{author}{M.~Prado~Rodriguez},
  \bibinfo{author}{P.~B. Price}, \bibinfo{author}{B.~Pries}, \bibinfo{author}{G.~T. Przybylski}, \bibinfo{author}{C.~Raab}, \bibinfo{author}{A.~Raissi}, \bibinfo{author}{M.~Rameez}, \bibinfo{author}{K.~Rawlins}, \bibinfo{author}{I.~C. Rea}, \bibinfo{author}{A.~Rehman}, \bibinfo{author}{P.~Reichherzer}, \bibinfo{author}{R.~Reimann}, \bibinfo{author}{G.~Renzi}, \bibinfo{author}{E.~Resconi}, \bibinfo{author}{S.~Reusch}, \bibinfo{author}{W.~Rhode}, \bibinfo{author}{M.~Richman}, \bibinfo{author}{B.~Riedel}, \bibinfo{author}{E.~J. Roberts}, \bibinfo{author}{S.~Robertson}, \bibinfo{author}{G.~Roellinghoff}, \bibinfo{author}{M.~Rongen}, \bibinfo{author}{C.~Rott}, \bibinfo{author}{T.~Ruhe}, \bibinfo{author}{D.~Ryckbosch}, \bibinfo{author}{D.~Rysewyk~Cantu}, \bibinfo{author}{I.~Safa}, \bibinfo{author}{J.~Saffer}, \bibinfo{author}{S.~E. Sanchez~Herrera}, \bibinfo{author}{A.~Sandrock}, \bibinfo{author}{J.~Sandroos}, \bibinfo{author}{M.~Santander}, \bibinfo{author}{S.~Sarkar}, \bibinfo{author}{S.~Sarkar},
  \bibinfo{author}{K.~Satalecka}, \bibinfo{author}{M.~Scharf}, \bibinfo{author}{M.~Schaufel}, \bibinfo{author}{H.~Schieler}, \bibinfo{author}{S.~Schindler}, \bibinfo{author}{P.~Schlunder}, \bibinfo{author}{T.~Schmidt}, \bibinfo{author}{A.~Schneider}, \bibinfo{author}{J.~Schneider}, \bibinfo{author}{F.~G. Schr{\"o}der}, \bibinfo{author}{L.~Schumacher}, \bibinfo{author}{G.~Schwefer}, \bibinfo{author}{S.~Sclafani}, \bibinfo{author}{D.~Seckel}, \bibinfo{author}{S.~Seunarine}, \bibinfo{author}{A.~Sharma}, \bibinfo{author}{S.~Shefali}, \bibinfo{author}{M.~Silva}, \bibinfo{author}{B.~Skrzypek}, \bibinfo{author}{B.~Smithers}, \bibinfo{author}{R.~Snihur}, \bibinfo{author}{J.~Soedingrekso}, \bibinfo{author}{D.~Soldin}, \bibinfo{author}{C.~Spannfellner}, \bibinfo{author}{G.~M. Spiczak}, \bibinfo{author}{C.~Spiering}, \bibinfo{author}{J.~Stachurska}, \bibinfo{author}{M.~Stamatikos}, \bibinfo{author}{T.~Stanev}, \bibinfo{author}{R.~Stein}, \bibinfo{author}{J.~Stettner}, \bibinfo{author}{A.~Steuer},
  \bibinfo{author}{T.~Stezelberger}, \bibinfo{author}{T.~St{\"u}rwald}, \bibinfo{author}{T.~Stuttard}, \bibinfo{author}{G.~W. Sullivan}, \bibinfo{author}{I.~Taboada}, \bibinfo{author}{F.~Tenholt}, \bibinfo{author}{S.~{Ter-Antonyan}}, \bibinfo{author}{S.~Tilav}, \bibinfo{author}{F.~Tischbein}, \bibinfo{author}{K.~Tollefson}, \bibinfo{author}{L.~Tomankova}, \bibinfo{author}{C.~T{\"o}nnis}, \bibinfo{author}{S.~Toscano}, \bibinfo{author}{D.~Tosi}, \bibinfo{author}{A.~Trettin}, \bibinfo{author}{M.~Tselengidou}, \bibinfo{author}{C.~F. Tung}, \bibinfo{author}{A.~Turcati}, \bibinfo{author}{R.~Turcotte}, \bibinfo{author}{C.~F. Turley}, \bibinfo{author}{J.~P. Twagirayezu}, \bibinfo{author}{B.~Ty}, \bibinfo{author}{M.~A. Unland~Elorrieta}, \bibinfo{author}{N.~{Valtonen-Mattila}}, \bibinfo{author}{J.~Vandenbroucke}, \bibinfo{author}{N.~{van Eijndhoven}}, \bibinfo{author}{D.~Vannerom}, \bibinfo{author}{J.~{van Santen}}, \bibinfo{author}{S.~Verpoest}, \bibinfo{author}{M.~Vraeghe}, \bibinfo{author}{C.~Walck},
  \bibinfo{author}{T.~B. Watson}, \bibinfo{author}{C.~Weaver}, \bibinfo{author}{P.~Weigel}, \bibinfo{author}{A.~Weindl}, \bibinfo{author}{M.~J. Weiss}, \bibinfo{author}{J.~Weldert}, \bibinfo{author}{C.~Wendt}, \bibinfo{author}{J.~Werthebach}, \bibinfo{author}{M.~Weyrauch}, \bibinfo{author}{N.~Whitehorn}, \bibinfo{author}{C.~H. Wiebusch}, \bibinfo{author}{D.~R. Williams}, \bibinfo{author}{M.~Wolf}, \bibinfo{author}{K.~Woschnagg}, \bibinfo{author}{G.~Wrede}, \bibinfo{author}{J.~Wulff}, \bibinfo{author}{X.~W. Xu}, \bibinfo{author}{Y.~Xu}, \bibinfo{author}{J.~P. Yanez}, \bibinfo{author}{S.~Yoshida}, \bibinfo{author}{S.~Yu}, \bibinfo{author}{T.~Yuan}, \bibinfo{author}{Z.~Zhang},
\newblock \bibinfo{title}{Combining {{Maximum-Likelihood}} with {{Deep Learning}} for {{Event Reconstruction}} in {{IceCube}}},
\newblock \bibinfo{journal}{Proc. Sci.} \bibinfo{volume}{395} (\bibinfo{year}{2022}) \bibinfo{pages}{1065}. \DOIprefix\doi{10.22323/1.395.1065}.
\bibitem[{Salinas et~al.(2020)Salinas, Flunkert, Gasthaus, and Januschowski}]{Salinas2020a}
\bibinfo{author}{D.~Salinas}, \bibinfo{author}{V.~Flunkert}, \bibinfo{author}{J.~Gasthaus}, \bibinfo{author}{T.~Januschowski},
\newblock \bibinfo{title}{{{DeepAR}}: {{Probabilistic}} forecasting with autoregressive recurrent networks},
\newblock \bibinfo{journal}{Int. J. Forecast.} \bibinfo{volume}{36} (\bibinfo{year}{2020}) \bibinfo{pages}{1181--1191}. \DOIprefix\doi{10.1016/J.IJFORECAST.2019.07.001}. \href{http://arxiv.org/abs/1704.04110}{{\tt arXiv:1704.04110}}.
\bibitem[{{Lloyd-Smith}(2007)}]{Lloyd-Smith2007a}
\bibinfo{author}{J.~O. {Lloyd-Smith}},
\newblock \bibinfo{title}{Maximum {{Likelihood Estimation}} of the {{Negative Binomial Dispersion Parameter}} for {{Highly Overdispersed Data}}, with {{Applications}} to {{Infectious Diseases}}},
\newblock \bibinfo{journal}{PLOS ONE} \bibinfo{volume}{2} (\bibinfo{year}{2007}) \bibinfo{pages}{e180}. \DOIprefix\doi{10.1371/JOURNAL.PONE.0000180}.
\bibitem[{Praszalowicz(2011)}]{Praszalowicz2011}
\bibinfo{author}{M.~Praszalowicz},
\newblock \bibinfo{title}{Negative {{Binomial Distribution}} and the multiplicity moments at the {{LHC}}},
\newblock \bibinfo{journal}{Phys. Lett. B} \bibinfo{volume}{704} (\bibinfo{year}{2011}) \bibinfo{pages}{566--569}. \DOIprefix\doi{10.1016/j.physletb.2011.09.101}.
\bibitem[{Tezlaf(2023)}]{Tezlaf2023}
\bibinfo{author}{S.~V. Tezlaf},
\newblock \bibinfo{title}{Significance of the negative binomial distribution in multiplicity phenomena},
\newblock \bibinfo{journal}{Phys. Scr.} \bibinfo{volume}{98} (\bibinfo{year}{2023}) \bibinfo{pages}{115310}. \DOIprefix\doi{10.1088/1402-4896/acfead}.
\bibitem[{Perez et~al.(2021)Perez, Malhotra, Rhoads, and Tilvi}]{Perez2021}
\bibinfo{author}{L.~A. Perez}, \bibinfo{author}{S.~Malhotra}, \bibinfo{author}{J.~E. Rhoads}, \bibinfo{author}{V.~Tilvi},
\newblock \bibinfo{title}{Void {{Probability Function}} of {{Simulated Surveys}} of {{High-redshift Ly$\alpha$ Emitters}}},
\newblock \bibinfo{journal}{ApJ} \bibinfo{volume}{906} (\bibinfo{year}{2021}) \bibinfo{pages}{58}. \DOIprefix\doi{10.3847/1538-4357/abc88b}.
\bibitem[{Fry and Colombi(2013)}]{Fry2013}
\bibinfo{author}{J.~N. Fry}, \bibinfo{author}{S.~Colombi},
\newblock \bibinfo{title}{Void statistics and hierarchical scaling in the halo model},
\newblock \bibinfo{journal}{MNRAS} \bibinfo{volume}{433} (\bibinfo{year}{2013}) \bibinfo{pages}{581--590}. \DOIprefix\doi{10.1093/mnras/stt745}.
\bibitem[{{Hurtado-Gil} et~al.(2017){Hurtado-Gil}, Mart{\'i}nez, {Arnalte-Mur}, {Pons-Border{\'i}a}, {Pareja-Flores}, and Paredes}]{Hurtado-Gil2017}
\bibinfo{author}{L.~{Hurtado-Gil}}, \bibinfo{author}{V.~J. Mart{\'i}nez}, \bibinfo{author}{P.~{Arnalte-Mur}}, \bibinfo{author}{M.-J. {Pons-Border{\'i}a}}, \bibinfo{author}{C.~{Pareja-Flores}}, \bibinfo{author}{S.~Paredes},
\newblock \bibinfo{title}{The best fit for the observed galaxy counts-in-cell distribution function},
\newblock \bibinfo{journal}{A\&A} \bibinfo{volume}{601} (\bibinfo{year}{2017}) \bibinfo{pages}{A40}. \DOIprefix\doi{10.1051/0004-6361/201629097}.
\bibitem[{Hameeda et~al.(2021)Hameeda, Plastino, and Rocca}]{Hameeda2021}
\bibinfo{author}{M.~Hameeda}, \bibinfo{author}{A.~Plastino}, \bibinfo{author}{M.~C. Rocca},
\newblock \bibinfo{title}{Generalized {{Poisson}} distributions for systems with two-particle interactions},
\newblock \bibinfo{journal}{IOPSciNotes} \bibinfo{volume}{2} (\bibinfo{year}{2021}) \bibinfo{pages}{015003}. \DOIprefix\doi{10.1088/2633-1357/abec9f}.
\bibitem[{Speagle(2020)}]{Speagle2020a}
\bibinfo{author}{J.~S. Speagle},
\newblock \bibinfo{title}{Dynesty: A dynamic nested sampling package for estimating {{Bayesian}} posteriors and evidences},
\newblock \bibinfo{journal}{MNRAS} \bibinfo{volume}{493} (\bibinfo{year}{2020}) \bibinfo{pages}{3132--3158}. \DOIprefix\doi{10.1093/mnras/staa278}.
\bibitem[{Kass and Raftery(1995)}]{Kass1995}
\bibinfo{author}{R.~E. Kass}, \bibinfo{author}{A.~E. Raftery},
\newblock \bibinfo{title}{Bayes {{Factors}}},
\newblock \bibinfo{journal}{J. Am. Stat. Assoc.} \bibinfo{volume}{90} (\bibinfo{year}{1995}) \bibinfo{pages}{773--795}. \DOIprefix\doi{10.1080/01621459.1995.10476572}.
\bibitem[{Skilling(2006)}]{Skilling2006a}
\bibinfo{author}{J.~Skilling},
\newblock \bibinfo{title}{Nested sampling for general {{Bayesian}} computation},
\newblock \bibinfo{journal}{Bayesian Anal.} \bibinfo{volume}{1} (\bibinfo{year}{2006}) \bibinfo{pages}{833--859}. \DOIprefix\doi{10.1214/06-BA127}.
\bibitem[{Feroz et~al.(2009)Feroz, Hobson, and Bridges}]{Feroz2009a}
\bibinfo{author}{F.~Feroz}, \bibinfo{author}{M.~P. Hobson}, \bibinfo{author}{M.~Bridges},
\newblock \bibinfo{title}{{{MultiNest}}: {{An}} efficient and robust {{Bayesian}} inference tool for cosmology and particle physics},
\newblock \bibinfo{journal}{Mon. Not. R. Astron. Soc.} \bibinfo{volume}{398} (\bibinfo{year}{2009}) \bibinfo{pages}{1601--1614}. \DOIprefix\doi{10.1111/J.1365-2966.2009.14548.X/2/M_MNRAS0398-1601-M32.GIF}. \href{http://arxiv.org/abs/0809.3437}{{\tt arXiv:0809.3437}}.
\bibitem[{Buchner(2021)}]{Buchner2021c}
\bibinfo{author}{J.~Buchner},
\newblock \bibinfo{title}{{{UltraNest}} - a robust, general purpose {{Bayesian}} inference engine},
\newblock \bibinfo{journal}{J. Open Source Softw.} \bibinfo{volume}{6} (\bibinfo{year}{2021}) \bibinfo{pages}{3001}. \DOIprefix\doi{10.21105/JOSS.03001}. \href{http://arxiv.org/abs/2101.09604}{{\tt arXiv:2101.09604}}.
\bibitem[{Ashton et~al.(2022)Ashton, Bernstein, Buchner, Chen, Cs{\'a}nyi, Fowlie, Feroz, Griffiths, Handley, Habeck, Higson, Hobson, Lasenby, Parkinson, P{\'a}rtay, Pitkin, Schneider, Speagle, South, Veitch, Wacker, Wales, and Yallup}]{Ashton2022a}
\bibinfo{author}{G.~Ashton}, \bibinfo{author}{N.~Bernstein}, \bibinfo{author}{J.~Buchner}, \bibinfo{author}{X.~Chen}, \bibinfo{author}{G.~Cs{\'a}nyi}, \bibinfo{author}{A.~Fowlie}, \bibinfo{author}{F.~Feroz}, \bibinfo{author}{M.~Griffiths}, \bibinfo{author}{W.~Handley}, \bibinfo{author}{M.~Habeck}, \bibinfo{author}{E.~Higson}, \bibinfo{author}{M.~Hobson}, \bibinfo{author}{A.~Lasenby}, \bibinfo{author}{D.~Parkinson}, \bibinfo{author}{L.~B. P{\'a}rtay}, \bibinfo{author}{M.~Pitkin}, \bibinfo{author}{D.~Schneider}, \bibinfo{author}{J.~S. Speagle}, \bibinfo{author}{L.~South}, \bibinfo{author}{J.~Veitch}, \bibinfo{author}{P.~Wacker}, \bibinfo{author}{D.~J. Wales}, \bibinfo{author}{D.~Yallup},
\newblock \bibinfo{title}{Nested sampling for physical scientists},
\newblock \bibinfo{journal}{Nat. Rev. Methods Primer 2022 21} \bibinfo{volume}{2} (\bibinfo{year}{2022}) \bibinfo{pages}{1--22}. \DOIprefix\doi{10.1038/s43586-022-00121-x}. \href{http://arxiv.org/abs/2205.15570}{{\tt arXiv:2205.15570}}.
\bibitem[{Nelson et~al.(2020)Nelson, Ford, Buchner, Cloutier, D{\'i}az, Faria, Hara, Rajpaul, and Rukdee}]{Nelson2020}
\bibinfo{author}{B.~E. Nelson}, \bibinfo{author}{E.~B. Ford}, \bibinfo{author}{J.~Buchner}, \bibinfo{author}{R.~Cloutier}, \bibinfo{author}{R.~F. D{\'i}az}, \bibinfo{author}{J.~P. Faria}, \bibinfo{author}{N.~C. Hara}, \bibinfo{author}{V.~M. Rajpaul}, \bibinfo{author}{S.~Rukdee},
\newblock \bibinfo{title}{Quantifying the {{Bayesian Evidence}} for a {{Planet}} in {{Radial Velocity Data}}},
\newblock \bibinfo{journal}{AJ} \bibinfo{volume}{159} (\bibinfo{year}{2020}) \bibinfo{pages}{73}. \DOIprefix\doi{10.3847/1538-3881/ab5190}.
\bibitem[{Perrakis et~al.(2014)Perrakis, Ntzoufras, and Tsionas}]{Perrakis2014}
\bibinfo{author}{K.~Perrakis}, \bibinfo{author}{I.~Ntzoufras}, \bibinfo{author}{E.~G. Tsionas},
\newblock \bibinfo{title}{On the use of marginal posteriors in marginal likelihood estimation via importance sampling},
\newblock \bibinfo{journal}{Comput. Stat. Data Anal.} \bibinfo{volume}{77} (\bibinfo{year}{2014}) \bibinfo{pages}{54--69}. \DOIprefix\doi{10.1016/j.csda.2014.03.004}.
\bibitem[{Metodiev et~al.(2024)Metodiev, {Perrot-Dock{\`e}s}, Ouadah, Irons, Latouche, and Raftery}]{Metodiev2024}
\bibinfo{author}{M.~Metodiev}, \bibinfo{author}{M.~{Perrot-Dock{\`e}s}}, \bibinfo{author}{S.~Ouadah}, \bibinfo{author}{N.~J. Irons}, \bibinfo{author}{P.~Latouche}, \bibinfo{author}{A.~E. Raftery},
\newblock \bibinfo{title}{Easily {{Computed Marginal Likelihoods}} from {{Posterior Simulation Using}} the {{THAMES Estimator}}},
\newblock \bibinfo{journal}{Bayesian Anal.} \bibinfo{volume}{-1} (\bibinfo{year}{2024}) \bibinfo{pages}{1--28}. \DOIprefix\doi{10.1214/24-BA1422}.
\bibitem[{Llorente et~al.(2023)Llorente, Martino, Delgado, and {L{\'o}pez-Santiago}}]{Llorente2023}
\bibinfo{author}{F.~Llorente}, \bibinfo{author}{L.~Martino}, \bibinfo{author}{D.~Delgado}, \bibinfo{author}{J.~{L{\'o}pez-Santiago}},
\newblock \bibinfo{title}{Marginal {{Likelihood Computation}} for {{Model Selection}} and {{Hypothesis Testing}}: {{An Extensive Review}}},
\newblock \bibinfo{journal}{SIAM Rev.} \bibinfo{volume}{65} (\bibinfo{year}{2023}) \bibinfo{pages}{3--58}. \DOIprefix\doi{10.1137/20M1310849}.
\bibitem[{Prettyman et~al.(2006)Prettyman, Hagerty, Elphic, Feldman, Lawrence, McKinney, and Vaniman}]{Prettyman2006a}
\bibinfo{author}{T.~H. Prettyman}, \bibinfo{author}{J.~J. Hagerty}, \bibinfo{author}{R.~C. Elphic}, \bibinfo{author}{W.~C. Feldman}, \bibinfo{author}{D.~J. Lawrence}, \bibinfo{author}{G.~W. McKinney}, \bibinfo{author}{D.~T. Vaniman},
\newblock \bibinfo{title}{Elemental composition of the lunar surface: {{Analysis}} of gamma ray spectroscopy data from {{Lunar Prospector}}},
\newblock \bibinfo{journal}{J. Geophys. Res. Planets} \bibinfo{volume}{111} (\bibinfo{year}{2006}). \DOIprefix\doi{10.1029/2005JE002656}.
\bibitem[{Prettyman et~al.(2011)Prettyman, Feldman, McSween, Dingler, Enemark, Patrick, Storms, Hendricks, Morgenthaler, Pitman, and Reedy}]{Prettyman2011a}
\bibinfo{author}{T.~H. Prettyman}, \bibinfo{author}{W.~C. Feldman}, \bibinfo{author}{H.~Y. McSween}, \bibinfo{author}{R.~D. Dingler}, \bibinfo{author}{D.~C. Enemark}, \bibinfo{author}{D.~E. Patrick}, \bibinfo{author}{S.~A. Storms}, \bibinfo{author}{J.~S. Hendricks}, \bibinfo{author}{J.~P. Morgenthaler}, \bibinfo{author}{K.~M. Pitman}, \bibinfo{author}{R.~C. Reedy},
\newblock \bibinfo{title}{Dawn's gamma ray and neutron detector},
\newblock \bibinfo{journal}{Space Sci. Rev.} \bibinfo{volume}{163} (\bibinfo{year}{2011}) \bibinfo{pages}{371--459}. \DOIprefix\doi{10.1007/s11214-011-9862-0}.
\bibitem[{Peplowski(2016)}]{Peplowski2016c}
\bibinfo{author}{P.~N. Peplowski},
\newblock \bibinfo{title}{The global elemental composition of 433 {{Eros}}: {{First}} results from the {{NEAR}} gamma-ray spectrometer orbital dataset},
\newblock \bibinfo{journal}{Planet. Space Sci.} \bibinfo{volume}{134} (\bibinfo{year}{2016}) \bibinfo{pages}{36--51}. \DOIprefix\doi{10.1016/j.pss.2016.10.006}.
\bibitem[{Breitenmoser et~al.(2025)Breitenmoser, Stabilini, Kasprzak, and Mayer}]{Breitenmoser2025m}
\bibinfo{author}{D.~Breitenmoser}, \bibinfo{author}{A.~Stabilini}, \bibinfo{author}{M.~M. Kasprzak}, \bibinfo{author}{S.~Mayer}, \bibinfo{title}{Quantitative mobile gamma-ray spectrometry through {{Bayesian}} inference}, \bibinfo{year}{2025}. \DOIprefix\doi{10.48550/arXiv.2512.18769}. \href{http://arxiv.org/abs/2512.18769}{{\tt arXiv:2512.18769}}.
\bibitem[{Breitenmoser et~al.(2022)Breitenmoser, Butterweck, Kasprzak, Yukihara, and Mayer}]{Breitenmoser2022}
\bibinfo{author}{D.~Breitenmoser}, \bibinfo{author}{G.~Butterweck}, \bibinfo{author}{M.~M. Kasprzak}, \bibinfo{author}{E.~G. Yukihara}, \bibinfo{author}{S.~Mayer},
\newblock \bibinfo{title}{Experimental and {{Simulated Spectral Gamma-Ray Response}} of a {{NaI}}({{Tl}}) {{Scintillation Detector}} used in {{Airborne Gamma-Ray Spectrometry}}},
\newblock \bibinfo{journal}{Adv. Geosci.} \bibinfo{volume}{57} (\bibinfo{year}{2022}) \bibinfo{pages}{89--107}. \DOIprefix\doi{10.5194/ADGEO-57-89-2022}.
\bibitem[{Pozzi et~al.(2012)Pozzi, Clarke, Walsh, Miller, Dolan, Flaska, Wieger, Enqvist, Padovani, Mattingly, Chichester, and Peerani}]{Pozzi2012}
\bibinfo{author}{S.~A. Pozzi}, \bibinfo{author}{S.~D. Clarke}, \bibinfo{author}{W.~J. Walsh}, \bibinfo{author}{E.~C. Miller}, \bibinfo{author}{J.~L. Dolan}, \bibinfo{author}{M.~Flaska}, \bibinfo{author}{B.~M. Wieger}, \bibinfo{author}{A.~Enqvist}, \bibinfo{author}{E.~Padovani}, \bibinfo{author}{J.~K. Mattingly}, \bibinfo{author}{D.~L. Chichester}, \bibinfo{author}{P.~Peerani},
\newblock \bibinfo{title}{{{MCNPX-PoliMi}} for nuclear nonproliferation applications},
\newblock \bibinfo{journal}{Nucl. Instrum. Methods Phys. Res. A} \bibinfo{volume}{694} (\bibinfo{year}{2012}) \bibinfo{pages}{119--125}. \DOIprefix\doi{10.1016/j.nima.2012.07.040}.
\bibitem[{Lang et~al.(2018)Lang, Pienaar, Hogenbirk, Masson, Nolte, Zimbal, R{\"o}ttger, Benabderrahmane, and Bruno}]{Lang2018}
\bibinfo{author}{R.~F. Lang}, \bibinfo{author}{J.~Pienaar}, \bibinfo{author}{E.~Hogenbirk}, \bibinfo{author}{D.~Masson}, \bibinfo{author}{R.~Nolte}, \bibinfo{author}{A.~Zimbal}, \bibinfo{author}{S.~R{\"o}ttger}, \bibinfo{author}{M.~L. Benabderrahmane}, \bibinfo{author}{G.~Bruno},
\newblock \bibinfo{title}{Characterization of a deuterium--deuterium plasma fusion neutron generator},
\newblock \bibinfo{journal}{Nuclear Instruments and Methods in Physics Research Section A: Accelerators, Spectrometers, Detectors and Associated Equipment} \bibinfo{volume}{879} (\bibinfo{year}{2018}) \bibinfo{pages}{31--38}. \DOIprefix\doi{10.1016/j.nima.2017.10.001}.
\bibitem[{Liskien and Paulsen(1973)}]{Liskien1973}
\bibinfo{author}{H.~Liskien}, \bibinfo{author}{A.~Paulsen},
\newblock \bibinfo{title}{Neutron production cross sections and energies for the reactions {{T}}({\emph{p,n}}){{3He}}, {{D}}({\emph{d,n}}){{3He}}, and {{T}}({\emph{d,n}}){{4He}}},
\newblock \bibinfo{journal}{Atomic Data and Nuclear Data Tables} \bibinfo{volume}{11} (\bibinfo{year}{1973}) \bibinfo{pages}{569--619}. \DOIprefix\doi{10.1016/S0092-640X(73)80081-6}.
\bibitem[{Lopez et~al.(2025)Lopez, Pakari, Ballard, Clarke, and Pozzi}]{Lopez2025}
\bibinfo{author}{R.~Lopez}, \bibinfo{author}{O.~Pakari}, \bibinfo{author}{C.~Ballard}, \bibinfo{author}{S.~Clarke}, \bibinfo{author}{S.~Pozzi},
\newblock \bibinfo{title}{A simulation pipeline for fast neutron imaging and spectroscopy using quantified detector attributes},
\newblock \bibinfo{journal}{Radiat. Meas.} \bibinfo{volume}{184} (\bibinfo{year}{2025}) \bibinfo{pages}{107440}. \DOIprefix\doi{10.1016/j.radmeas.2025.107440}.
\bibitem[{McConn et~al.(2011)McConn, Gesh, Pagh, and Rucker}]{McConn2011a}
\bibinfo{author}{R.~J. McConn}, \bibinfo{author}{C.~J. Gesh}, \bibinfo{author}{R.~T. Pagh}, \bibinfo{author}{R.~A. Rucker}, \bibinfo{title}{Compendium of {{Material Composition Data}} for {{Radiation Transport Modeling}}}, \bibinfo{type}{Technical Report}, Pacific Northwest National Laboratory, \bibinfo{address}{Richland}, \bibinfo{year}{2011}.
\bibitem[{Jeffreys(1948)}]{Jeffreys1948}
\bibinfo{author}{H.~Jeffreys}, \bibinfo{title}{Theory {{Of Probability}}}, \bibinfo{edition}{2} ed., \bibinfo{publisher}{Oxford University Press}, \bibinfo{year}{1948}.
\bibitem[{Sellke et~al.(2001)Sellke, Bayarri, and Berger}]{Sellke2001}
\bibinfo{author}{T.~Sellke}, \bibinfo{author}{M.~J. Bayarri}, \bibinfo{author}{J.~O. Berger},
\newblock \bibinfo{title}{Calibration of {$\rho$} {{Values}} for {{Testing Precise Null Hypotheses}}},
\newblock \bibinfo{journal}{Am. Stat.} \bibinfo{volume}{55} (\bibinfo{year}{2001}) \bibinfo{pages}{62--71}. \DOIprefix\doi{10.1198/000313001300339950}.
\bibitem[{Cecil and Nieschmidt(1986)}]{Cecil1986}
\bibinfo{author}{F.~E. Cecil}, \bibinfo{author}{E.~B. Nieschmidt},
\newblock \bibinfo{title}{Production of 14 {{MeV}} neutrons from {{D-D}} neutron generators},
\newblock \bibinfo{journal}{Nuclear Instruments and Methods in Physics Research Section B: Beam Interactions with Materials and Atoms} \bibinfo{volume}{16} (\bibinfo{year}{1986}) \bibinfo{pages}{88--90}. \DOIprefix\doi{10.1016/0168-583X(86)90230-2}.
\bibitem[{{Foreman-Mackey}(2016)}]{Foreman-Mackey2016}
\bibinfo{author}{D.~{Foreman-Mackey}},
\newblock \bibinfo{title}{Corner.py: {{Scatterplot}} matrices in {{Python}}},
\newblock \bibinfo{journal}{J. Open Source Softw.} \bibinfo{volume}{1} (\bibinfo{year}{2016}) \bibinfo{pages}{24}. \DOIprefix\doi{10.21105/joss.00024}.
\bibitem[{Breitenmoser(2026)}]{Breitenmoser2026c}
\bibinfo{author}{D.~Breitenmoser},
\newblock \bibinfo{title}{Disentangling mixed neutron fields: Multi-source identification from few detected events ({{Dataset}})}  (\bibinfo{year}{2026}). \DOIprefix\doi{10.5281/zenodo.21516447}.

\end{thebibliography}

\end{document}